%% file: main-current.tex
\documentclass[letterpaper,superscriptaddress,twocolumn,aps,nofootinbib]{revtex4-2}

\usepackage{color}
\usepackage{amssymb,bm}
\usepackage{amsthm}
\usepackage{amsmath}
\usepackage{dsfont}
\usepackage[dvipsnames]{xcolor}
\usepackage{bbm}
\usepackage{xifthen}
\usepackage{graphicx}
\usepackage{nicefrac}
\usepackage{multirow}
\usepackage{booktabs}
\usepackage[normalem]{ulem}
\usepackage{cancel}
\usepackage{tabularx}
\usepackage{makecell}
\usepackage{mathtools}
\usepackage{soul}
\usepackage[caption=false]{subfig}

\usepackage[T1]{fontenc}
\usepackage[utf8]{inputenc}
\usepackage[colorlinks]{hyperref}
\hypersetup{colorlinks=true,linkcolor=red,anchorcolor=blue,citecolor=blue,urlcolor=blue}
\input{math_commands}

\AtBeginDocument{
\heavyrulewidth=.08em
\lightrulewidth=.05em
\cmidrulewidth=.03em
\belowrulesep=.65ex
\belowbottomsep=0pt
\aboverulesep=.4ex
\abovetopsep=0pt
\cmidrulesep=\doublerulesep
\cmidrulekern=.5em
\defaultaddspace=.5em}

\renewcommand{\eqref}[1]{Eq.~(\ref{#1})} 

\begin{document}
\title{Predicting Properties of Quantum Systems with Conditional Generative Models}

\author{Haoxiang Wang}
\thanks{Equal contribution (alphabetic order).}
\affiliation{University of Illinois Urbana-Champaign, Urbana, IL 61801, USA}
\affiliation{AWS Quantum Technologies, Seattle, WA 98170, USA}
\author{Maurice Weber}
\thanks{Equal contribution (alphabetic order).}
\affiliation{ETH Zürich, Department of Computer Science, 8092 Zürich, Switzerland}
\affiliation{Xanadu, Toronto, ON, M5G 2C8, Canada}
\author{Josh Izaac}
\affiliation{Xanadu, Toronto, ON, M5G 2C8, Canada}
\author{Cedric Yen-Yu Lin} 
\affiliation{AWS Quantum Technologies, Seattle, WA 98170, USA}

\begin{abstract}

Machine learning has emerged recently as a powerful tool for predicting properties of quantum many-body systems. For many ground states of gapped Hamiltonians, generative models can learn from measurements of a single quantum state to reconstruct the state accurately enough to predict local observables. 
Alternatively, classification and regression models can predict local observables by learning from measurements on different but related states. In this work, we combine the benefits of both approaches and propose the use of conditional generative models to simultaneously represent a family of states, learning shared structures of different quantum states from measurements. The trained model enables us to predict arbitrary local properties of ground states, even for states not included in the training data, without necessitating further training for new observables. 
We first numerically validate our approach on 2D random Heisenberg models using simulations of up to 45 qubits. Furthermore, we conduct quantum simulations on a neutral-atom quantum computer and demonstrate that our method can accurately predict the quantum phases of square lattices of 13$\times$13 Rydberg atoms.

\end{abstract}

\maketitle
\section{Introduction}
By harnessing the power of quantum mechanics, quantum computing has the potential to facilitate scientific discoveries and tackle various technological challenges. One of the most promising applications of quantum computing is to solve quantum many-body problems, a cornerstone task in multiple scientific disciplines, including physics, chemistry, and materials science~\cite{alexeev2021quantum}.
While classical methods such as density functional theory~\cite{burke2012perspective}, density-matrix renormalization group (DMRG)~\cite{white1992density}, and quantum Monte Carlo~\cite{ceperley1986quantum} have brought significant advancements in the study of certain systems of interest, they typically do not scale to large system sizes for general quantum mechanical systems. 
In contrast, due to their intrinsically quantum mechanical nature, quantum computers are expected to be able to simulate large-scale quantum many-body systems in the near future~\cite{alexeev2021quantum,altman2021quantum}.
The utility of such approaches crucially hinges on the ability to extract useful information from the simulated system by means of measuring multiple identical copies and subsequent statistical post-processing of the obtained measurements.
The task of complete characterization of the quantum state of interest is known as quantum state tomography (QST)~\cite{vogel1989determination,dadriano2003quantum,roos2004bell,gross2010quantum}. However, for generic systems, exact tomographic techniques scale unfavorably with system size and become impractical as the number of qubits increases~\cite{haffner2005scalable,lu2007experimental}. In cases where one is interested only in a limited number of observables of the state, shadow tomography~\cite{aaronson2019shadow,huang2020predicting,huang2021derandomization,hu2021classical,akhtar2022scalable,bertoni2022shallow} is a promising alternative to full-state tomography, alleviating the requirement to prepare an exponential number of copies of the state. Shadow tomography nevertheless has its limitations (on modern NISQ machines), e.g., in the case of global observables~\cite{huang2020predicting}.

Machine learning (ML) has emerged as a powerful and general technique for  QST in regimes where a limited amount of experimental data is available~\cite{torlai2018neural,carrasquilla2019reconstructing,rocchetto2018learning,ahmed2021quantum}. In this context, restricted Boltzmann machines (RBM) have been studied extensively and were shown to be able to learn accurate representations of states whose evolution can be simulated in polynomial time using classical methods~\cite{torlai2018neural}. The representational power of deep Boltzmann machines (DBM) with complex weights has been studied theoretically in Ref.~\cite{gao2017efficient}, where it has been proven that these approaches can efficiently represent most quantum states. While theoretically encouraging, DBMs with complex weights remain of limited practical value, due to the intrinsic difficulties of efficiently sampling from such models.
In response, other deep architectures have been explored, including recurrent neural networks (RNN)~\cite{carrasquilla2017machine}, variational autoencoders~\cite{rocchetto2018learning} and transformer networks~\cite{cha2021attention}.
These results point to the encouraging observation that these proposed neural networks have sufficient representational power to learn some complex quantum states. However, when the task is to learn representations of \emph{a family of quantum states}, these neural networks have to be re-trained on each state of the family to encode the state.

In an orthogonal line of research, ML has been used to directly predict phases of matter of quantum systems~\cite{carrasquilla2017machine, vargas2018extrapolating, miles2021machine}. These ML models are trained on measurement outcomes from many related quantum systems, and then used to predict phases of matter for unseen quantum systems from the same family; in contrast to QST, these techniques do not aim to learn a representation of the quantum states of interest. Huang et al.~\cite{huang2021provably} take this a step further to estimate arbitrary local (i.e., few-body) observables, by combining a kernel-based learning approach with classical shadows.
Specifically, given a collection of classical shadows obtained from performing random Pauli measurements on many different but related quantum states, this technique allows predicting the expectation of any local observable. 
In addition, Huang et al.~\cite{huang2021provably} provide rigorous guarantees for the classical representation obtained via kernel methods. As for numerical experiments, the authors of \cite{huang2021provably} fit an individual model for each observable, separately. Notably, while kernel methods enable rigorous guarantees, they usually empirically underperform neural networks in modern ML tasks \cite{arora2019exact,shankar2020neural}.

Here, we build on these previous works, and propose a framework for learning classical representations of a family of quantum states using machine learning (ML) models.
To that end, we propose an ML-based method to post-process measurement statistics of multiple quantum states \emph{jointly} and exploit the shared structure between these states, such as, e.g., ground states of a family of quantum many-body Hamiltonians.
This reduces the experimental and computational burden from at least two angles. Firstly, similar to Ref.~\cite{huang2021provably}, our technique enables experimentalists and practitioners to reduce measurement costs and even study properties of states for which no measurements are available, including states that cannot be prepared on modern NISQ hardware. However, in contrast to Ref.~\cite{huang2021provably}, our technique learns a representation of the family of states and allows us to generate new samples.
Secondly, similar to ML-based QST \cite{carrasquilla2019reconstructing,torlai2018neural,rocchetto2018learning}, we encode the full state in a generative model that allows us to predict arbitrary properties of the states by sampling from the model, alleviating the need to train a model for each property from scratch. In contrast to ML-based QST, our model is not restricted to a single state, but also generalizes to related states for which no measurements have been obtained.

In numerical simulations (of up to 45 qubits) of the 2D anti-ferromagnetic random Heisenberg model (Sec.~\ref{sec:heisenberg}), we show that our conditional generative models are able to accurately represent new ground states and predict their properties such as subsystem entanglement entropy and correlation functions.

Moreover, we conduct quantum simulations on Aquila \cite{aquila}, a 256-qubit neutral-atom quantum computer hosted on Amazon Braket \cite{braket}, to simulate a 2D square lattice system of Rydberg atoms. Specifically, we prepare different ground states of 13$\times$13 Rydberg atoms through adiabatic evolution, following the recipe of Ref.~\cite{Rydberg2D-256atoms}, and collect $Z$-basis measurement results of all atoms. We train our conditional generative model on a subset of the measurement data, showing that the trained model can accurately predict quantum phases of this 13$\times$13 Rydberg-atom system (Sec.\ref{sec:rydberg}). Furthermore, with proof-of-concept numerical examples (Sec.~\ref{sec:rydberg:larger-systems}, \ref{sec:rydberg:longer-evolution}), we demonstrate that our model can potentially enable experimentalists and practitioners to predict properties of systems beyond current hardware limits, such as large-scale quantum systems that modern quantum computers cannot simulate.

\section{Representing quantum states with conditional generative models}

\begin{figure}[t]
    \centering
    \includegraphics[width=\linewidth]{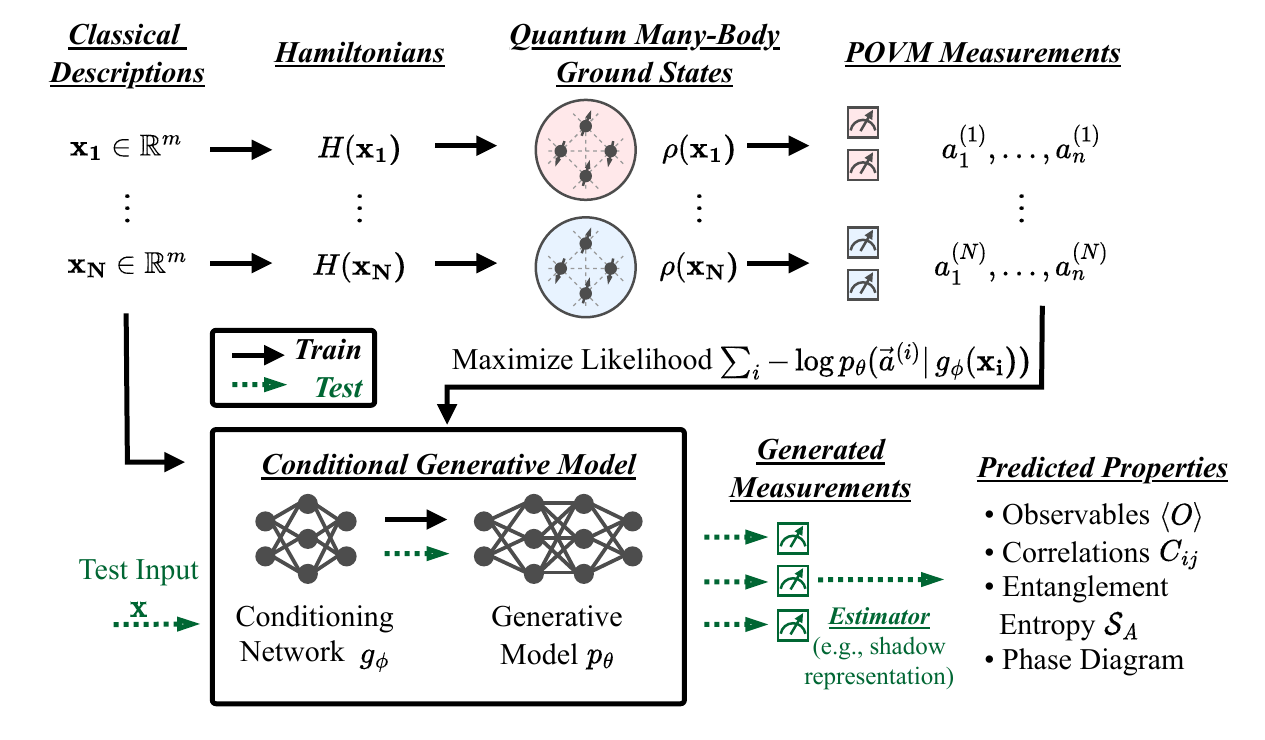}
    \caption{Overview of our framework. The training pipeline of the conditional generative model is marked with solid black arrows, while the test pipeline is marked with dashed green arrows.}
    \label{fig:framework}
\end{figure}

In this work, we consider $n$-qubit quantum states $\rho(\mathbf{x})$ to be density operators acting on a $d$-dimensional complex Hilbert space $\cH$, with $d=2^n$. The density operators are parametrized by a real-valued parameter $\mathbf{x}\in\R^m$. The most general type of measurement allowed by quantum mechanics is positive operator-valued measures (POVM). A POVM is a collection of positive semidefinite linear operators $\cM = \{M_{\vec a}\}_{\vec a}$ which resolve the identity $\sum_{\vec a}M_{\vec a} = \Id$. Each POVM operator $M_{\vec a}$ corresponds to a measurement outcome $\vec a = (a_1,\,\ldots,\,a_n)$ whose probability is obtained from Born's rule,
\begin{equation}
    p(\vec a\,\lvert\,\mathbf{x}) = \tr{M_{\vec a} \rho(\mathbf{x})}.
\end{equation}
To fully represent a generic quantum state $\rho$, we need to use an informationally complete (IC) POVM; that is, the POVM must consist of at least $d^2$ elements which span the space of self-adjoint operators on $\cH$. This implies that the measurements collected from an IC POVM contain the necessary statistical information to fully characterize the state.
In this work, in the context of the Heisenberg model, we focus on the Pauli-6 POVM consisting of $n$-qubit tensor products of projections to the eigenspaces of the Pauli observables~\cite{carrasquilla2019reconstructing}\footnote{We refer the reader to Appendix~\ref{apx:simulation:heisenberg} for details about the Pauli-6 POVM.}.
This POVM corresponds to performing measurements in one of the three Pauli bases, uniformly at random. 
For the Rydberg atom systems, we make measurements in the computational basis (i.e., $Z$-basis) as this POVM is sufficient for characterizing ground states of the Rydberg systems considered here.
We remark that our proposed framework is not restricted to these choices of POVMs (other POVMs can also be adopted). 
Among different ways to estimate properties like expectation values or entanglement entropies, we adopt the classical shadow protocol~\cite{huang2020predicting} due to its simplicity and statistical guarantees.

For a family of states $\rho(\mathbf{x})$ indexed by the parameter $\mathbf{x}$, our dataset consists of a collection of measurement statistics and parameters, $\cD = \{(\vec a^{(1)},\,\mathbf{x}^{(1)}),\,\ldots,\,(\vec a^{(N)},\,\mathbf{x}^{(N)})\}$, and we train a generative model $p_{\theta}$, parameterized as a neural network, to maximize the likelihood of the observed measurements. To condition the model distribution $p_{\theta}$ on the classical parameter $\mathbf{x}$, we use an embedding $\mathbf{x} \mapsto g_\phi(\mathbf{x})$ with trainable parameters $\phi$ and use this as an additional input to the model $p_{\theta}$. To make the optimization and sampling process tractable, we decompose the joint distribution into a product of conditional distributions in an autoregressive manner,
\begin{equation}
    \label{eq:model-distribution}
    p_{\theta,\phi}(a_1,\,\ldots,\,a_n\,\lvert\,\mathbf{x}) = \prod_{i=1}^n p_\theta(a_i\,\lvert\,a_{i-1},\,\ldots,\,a_1,\,g_\phi(\mathbf{x})),
\end{equation}
where $\theta$ denotes the set of parameters of the generative model, and $\phi$ are the parameters of the embedding network. Since both the embedding $g_\phi$ and the generative model $p_\theta$ are parametrized as neural networks, they can be trained in an end-to-end manner. 

\textbf{Training.}~To train the model, we sample $N_H$ Hamiltonians (each of a unique parameter $\mathbf{x}\in\R^m$) and obtain $N_s$ measurement outcomes for each ground state $\rho(\mathbf{x})$. Each measurement outcome is a training sample, leading to a training set $\cD$ of size $N_H N_s$. As our training objective, we minimize the average negative log-likelihood loss over training data,
\begin{align}
    \min_{\theta,\phi} \cL(\theta,\,\phi) \coloneqq \frac{1}{N_H N_s}\sum_{(\vec a,\mathbf{x})\in \cD}{-\log p_{\theta,\phi}(a_1,\,\ldots,\,a_n~\lvert~ \mathbf{x})},
\end{align}
what corresponds to maximizing the (conditional) likelihoods over the observed measurement outcomes.

\textbf{Prediction.}~Once the model is trained, measurements for a new state of the same family (e.g., ground state of a related Hamiltonian) can be generated by conditioning it on the classical variable. Based on these generated measurements, arbitrary properties of the unknown state can be predicted using estimation protocols such as shadow tomography~\cite{huang2020predicting}. A schematic diagram of our framework is presented in Figure~\ref{fig:framework}.

\section{Learning ground states of quantum many-body Hamiltonians}
In this section, we present our results for two quantum many-body problems, 2D random Heisenberg models and Rydberg atom systems, where each problem includes a family of Hamiltonians. Our model can learn from ground-state measurements of multiple Hamiltonians, and be used to predict ground-state properties of other Hamiltonians from the same family.
For 2D random Heisenberg models, our proposed conditional generative model (CGM) consists of i) a graph convolutional network (GCN) \cite{kipf2017semisupervised} which takes the classical description of the Hamiltonian as input, and ii) a transformer \cite{vaswani2017attention} as the generative model to represent quantum states and generate measurements. For Rydberg atom systems, we use a linear projection instead of a GCN to condition on system parameters, and continue to use a transformer as the generative model. In both cases, our implementations are publicly available\footnote{\url{https://github.com/PennyLaneAI/generative-quantum-states}}.

\section{Problem I: 2D anti-ferromagnetic random Heisenberg model}\label{sec:heisenberg}
The first family of Hamiltonians we consider is the 2D anti-ferromagnetic random Heisenberg model, where qubits, i.e., spin-1/2 particles, are allocated over a square lattice. Specifically, we are interested in the ground state of the Hamiltonians
\begin{align}\label{eq:H-heisenberg}
    H(\bx) = \sum_{\langle ij \rangle} \mathbf{x}_{ij} (X_i X_j + Y_i Y_j + Z_i Z_j),
\end{align}
where $\langle ij \rangle$ represents nearest-neighbor interactions and the summation is over all possible pairs on this lattice. For each pair $\langle ij \rangle$, the corresponding interaction strength $\bx_{ij}$ is uniformly sampled from the interval $[0,2]$. The Hamiltonian in~\eqref{eq:H-heisenberg} can be described by a weighted undirected graph without any loss of information. Each qubit is expressed as a node, and the coupling strength between two sites corresponds to a weighted edge in the graph. Denoting the adjacency matrix of the graph as $\mathbf{x}$, we write $\rho(\mathbf{x})$ for the ground states of the Hamiltonian corresponding to the grid lattice defined by the graph $\mathbf{x}$.

\textbf{Model Structure.}~To exploit the full classical information contained in the graph structure of the Hamiltonian, we use a graph convolutional network (GCN)~\cite{kipf2017semisupervised}, followed by a linear projection layer to get a trainable embedding of the graph, $\mathbf{x} \mapsto g_\phi(\mathbf{x})$.
As for the generative model, we adopt the transformer architecture~\cite{vaswani2017attention}, which has shown tremendous success for sequence modeling in the context of natural language processing~\cite{brown2020language,devlin2018bert}, computer vision~\cite{liu2021swin} and also for quantum state tomography~\cite{cha2021attention}. We condition the transformer on the Hamiltonian by adding the graph embedding vector $g_\phi(\mathbf{x})$ to the encoded measurements after the positional encoding. 
The model structure is illustrated in Figure~\ref{fig:heisenberg-generative-model}, and more details are provided in Appendix~\ref{apx:transformer-architecture}.

\textbf{Estimation Protocol.}~Given a trained model for a fixed lattice size, we evaluate it by sampling from the model distribution~
(\ref{eq:model-distribution}) and estimate properties using the classical shadow protocol~\cite{huang2020predicting}.
Specifically, given a collection of $N$ artificial measurements outcomes generated by the model, $\Vec{b}^{(1)},\,\ldots,\,\Vec{b}^{(N)}$, the resulting shadow state can be expressed as
\begin{equation}
    \label{eq:model-shadow-state}
    \hat\rho(\mathbf{x})_{\mathrm{model}} = \frac{1}{N}\sum_{i=1}^N \bigotimes_{j=1}^n\left(3\ketbra{b_j^{(i)}}{b_j^{(i)}} - \Id_2\right)
\end{equation}
where the kets $\ket{b_j^{(i)}}$ correspond to the eigenstate associated with the measurement outcome $b_j^{(i)}$ sampled by the model (e.g., if $+$ was sampled, then $\ket{b_j^{(i)}} = \ket{+}$).
Based on this representation of the state, it is then straightforward to evaluate properties such as expectation values of observables.
Throughout the experiments in this section, we fix the number of samples generated by our generative models to $N=20000$.

\begin{figure}
    \centering
    \includegraphics[width=\linewidth]{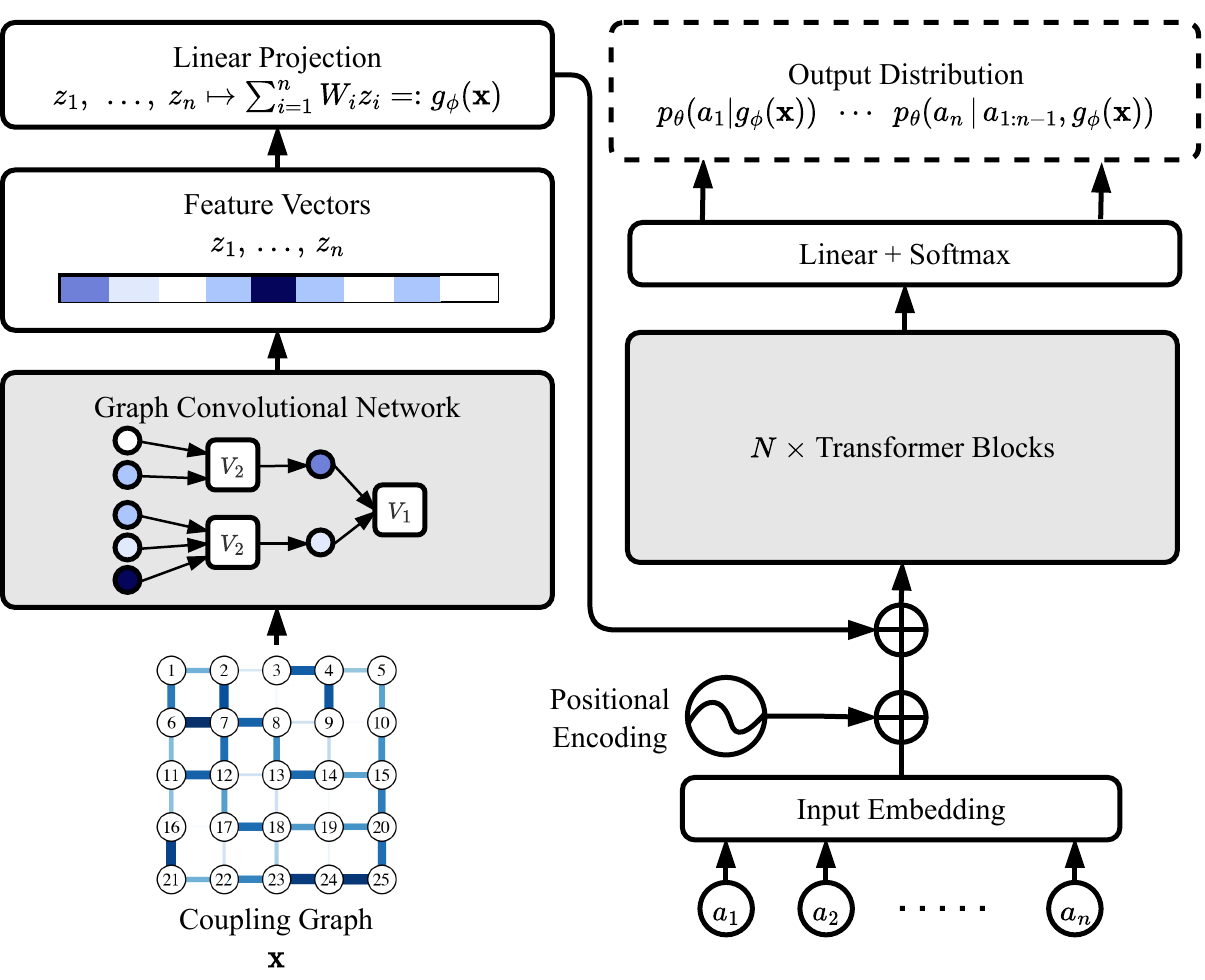}
    \caption{Overview of the conditional generative model used for ground states of the 2D anti-ferromagnetic Heisenberg model.}
    \label{fig:heisenberg-generative-model}
\end{figure}

\textbf{Dataset.}~The dataset for training and evaluation is obtained from classical simulations. For systems with fewer than 20 qubits, and for a lattice structure with $n_r$ rows and $n_c$ columns, we generate 100 random Hamiltonians by sampling coupling constants uniformly at random $J_{ij}\overset{\mathrm{iid}}{\sim}\cU[0,\,2]$, and determine the ground state via exact diagonalization. For systems with 20 qubits or more, we use the publicly available dataset\footnote{The data is available at \url{https://github.com/hsinyuan-huang/provable-ml-quantum}} provided by the authors of Ref.~\cite{huang2021provably} which was obtained via DMRG~\cite{white1992density}. We measure the Pauli-6 POVM, which effectively corresponds to measuring the state in random Pauli bases.

\textbf{Training.}~For each randomly sampled Hamiltonian, we use 1000 measurements for the ground state of each Hamiltonian, resulting in a data set with 100000 randomized Pauli measurements. We train our model on 80 Hamiltonians, and set aside the remaining 20 Hamiltonians as the test set to evaluate how well our models generalize to new lattice structures. In the training stage, we adopt common techniques for transformer training, such as Adam optimizer~\cite{adam}, layer normalization~\cite{layernorm}, and cosine learning-rate annealing~\cite{loshchilov2017sgdr}. 
We refer the reader to Appendix~\ref{apx:exp-details:heisenberg} for details about the training procedure.

\subsection{Predicting correlation functions}

\begin{figure*}
    \centering
    \subfloat[Coupling graph]{\label{subfig:heisenberg-correlations-couplings}\includegraphics[width=.22\linewidth]{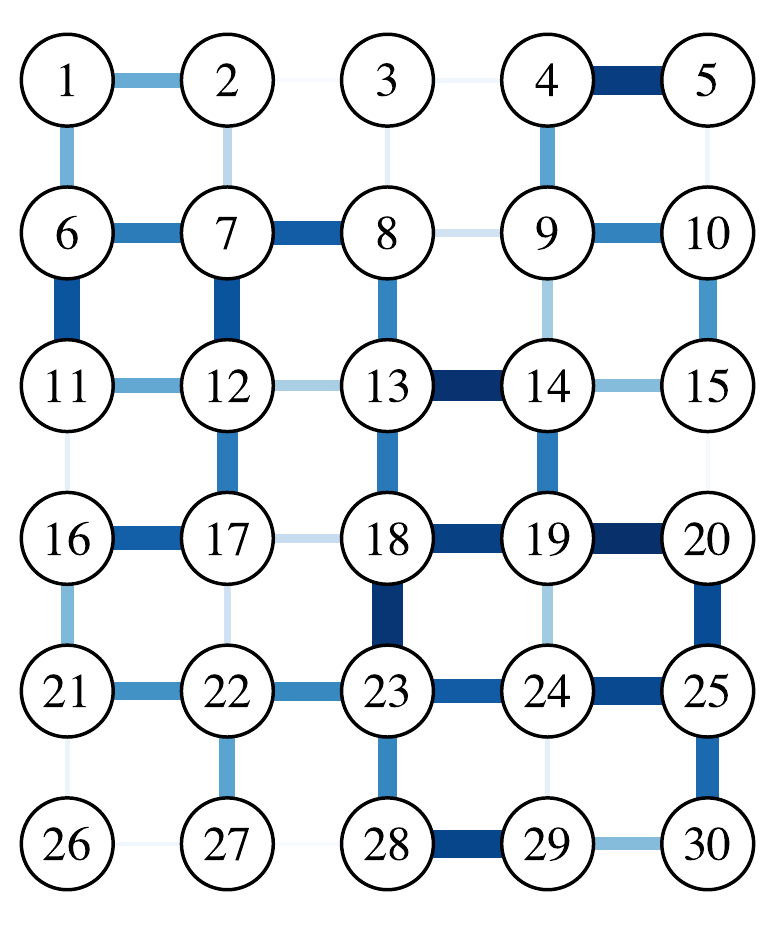}}%
    \hfill
    \subfloat[Two-point correlation functions]{\label{subfig:heisenberg-correlations}\includegraphics[width=.55\linewidth]{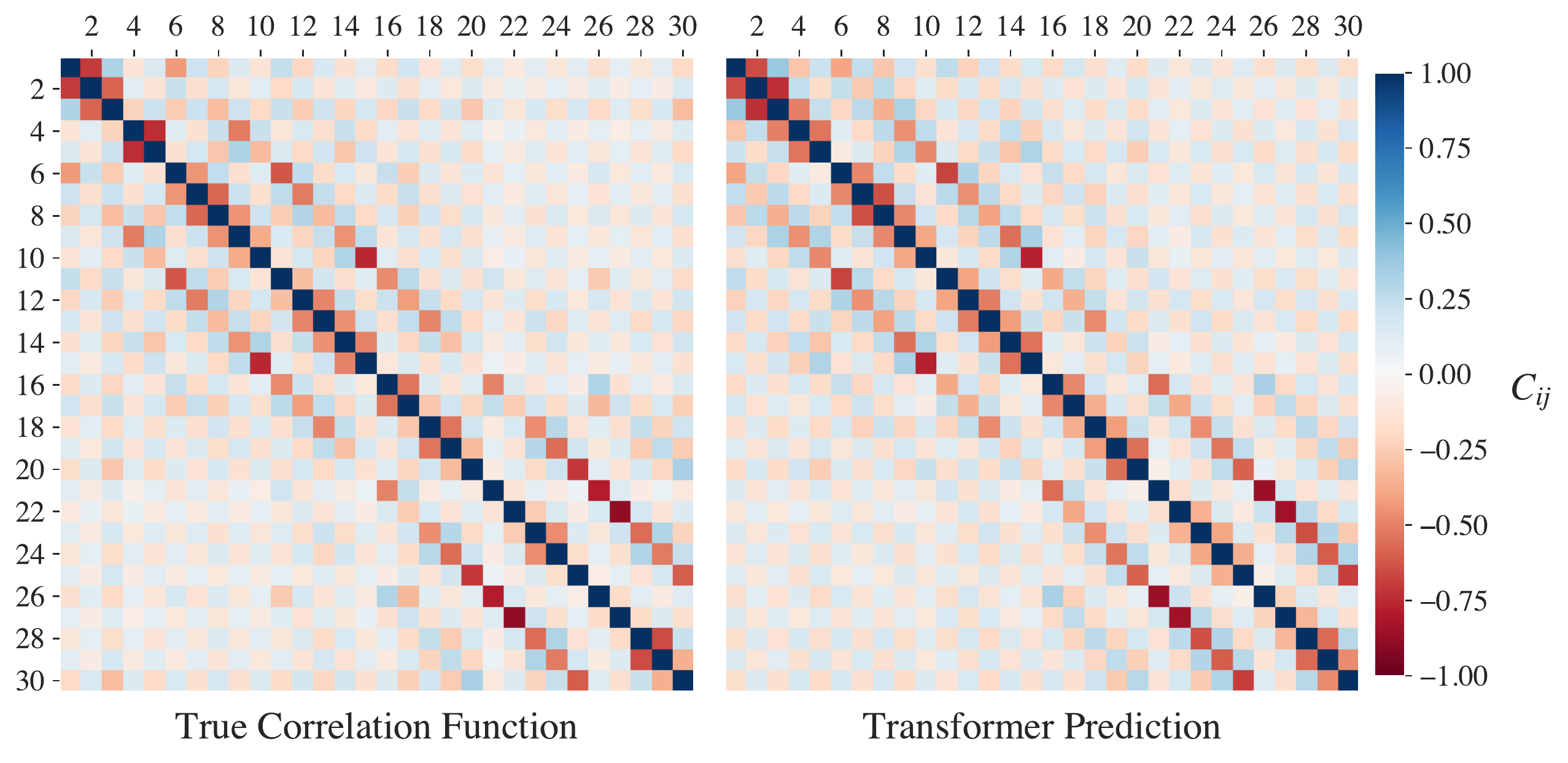}}%
    \hfill
    \subfloat[Correlation functions RMSE]{\label{subfig:heisenberg-correlations-rmse}\includegraphics[width=.21\linewidth]{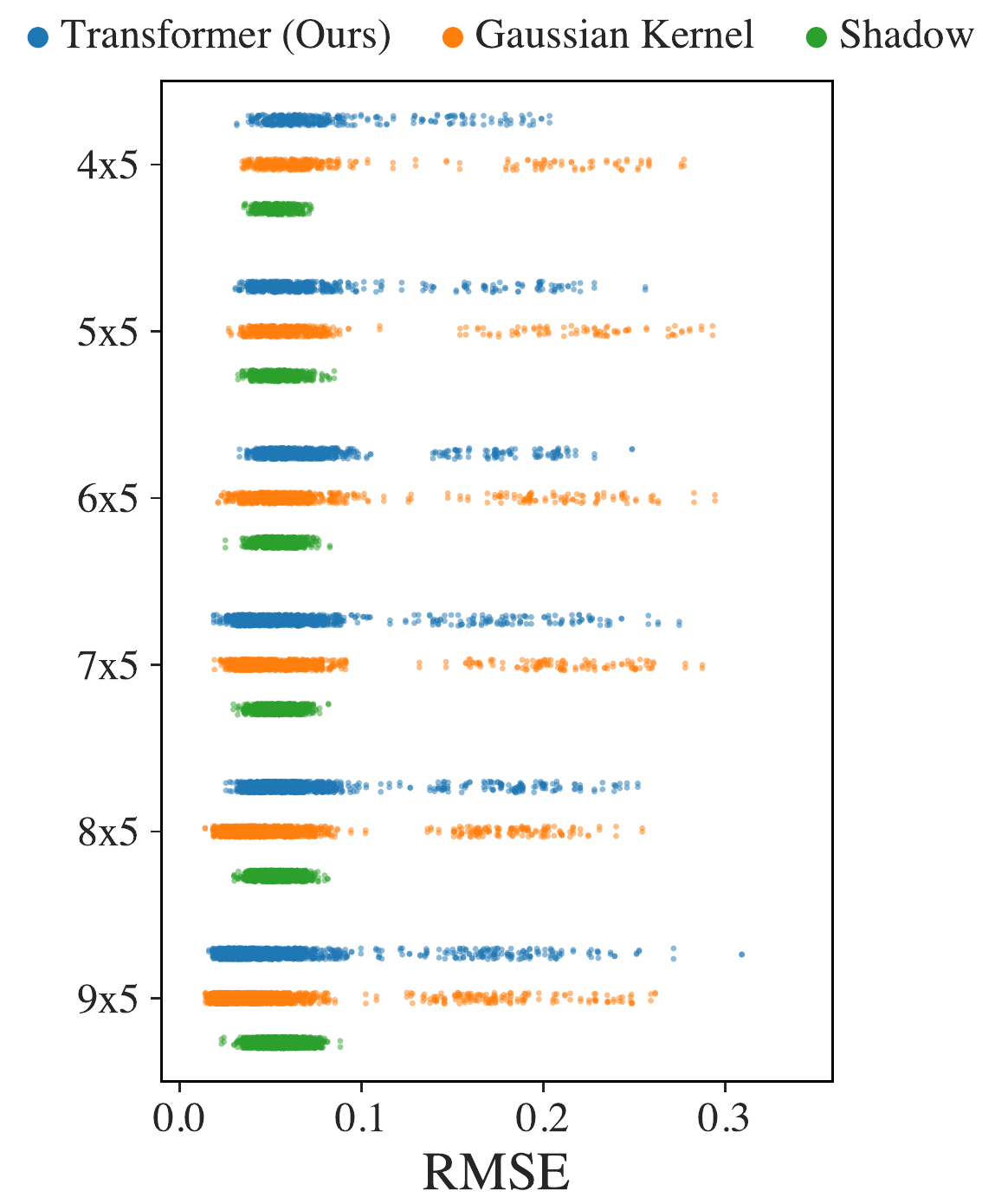}}
    \caption{Predicting correlation functions of ground states of the 2D random anti-ferromagnetic Heisenberg model. a) Random Coupling graph from the test set. The graph determines the 2D random Heisenberg model~(\ref{eq:H-heisenberg}), and is used to condition our generative model. The interaction strength is indicated by both thickness and color (thicker and darker means higher interaction strength). b) True and predicted two-point correlation functions~(\ref{eq:correlation-functions}) for a ground state from the test set encoded by our generative model given the coupling graph.
    c) Root Mean Square Error (RMSE) between true and predicted correlation functions for systems of different sizes, for our conditional generative model (blue), Gaussian kernel (orange), and shadow tomography (green). Each point in the plot corresponds to the error of correlation predictions for different sites, averaged over Hamiltonians from the test set.}
\end{figure*}

We first consider two-point correlation functions between two sites $(i,\,j)\mapsto \langle C_{ij}\rangle$ with
\begin{equation}
    \label{eq:correlation-functions}
    C_{ij} = \frac{1}{3}\left(X_iX_j +Y_iY_j + Z_iZ_j\right).
\end{equation}
Fig.~\ref{subfig:heisenberg-correlations} shows the true and predicted correlation functions for the Hamiltonian defined by the (randomly sampled) coupling graph shown in Fig.~\ref{subfig:heisenberg-correlations-couplings}. The predictions are obtained by conditioning the generative transformer on the coupling graph, and then generating new samples (i.e., POVM outcomes) to reconstruct a classical shadow. Based on the obtained classical shadow, we estimate the observables from~\eqref{eq:correlation-functions} for the correlation functions.
As can be seen from the figure, the model is able to encode a state for which the correlation function is fairly accurate and is close to the true correlation function.
In Fig.~\ref{subfig:heisenberg-correlations-rmse} we show the root mean square errors (RMSE) between predicted and true correlation functions for different system sizes and different techniques. Each point in the figure corresponds to the prediction error for particular sites $\{i,\,j\}$, averaged over all Hamiltonians in the test set. We also include the prediction errors for the Gaussian kernel method proposed in Ref.~\cite{huang2021provably}, which has been specifically trained to predict the function $\mathbf{x} \mapsto \langle C_{ij}\rangle_{\rho(\mathbf{x})}$ for a particular pair $\{i,\,j\}$ (i.e., a new model needs to be trained for each pair of sites $\{i,\,j\}$). Furthermore, we also include the error for the property predicted using classical shadows, which relies on $1000$ measurements.

\begin{itemize}
    \item \textbf{Comparison to the Gaussian kernel method}: It can be seen that our method performs similarly to the kernel method. We remark that our method is designed to solve a much more general task (i.e., encoding the entire family of states), while the Gaussian kernel only learns the --- arguably simpler --- classical-to-classical mapping $\mathbf{x} \mapsto \langle C_{ij}\rangle_{\rho(\mathbf{x})}$ for each pair of sites individually, and cannot be used to predict other properties.
    \item \textbf{Comparison to classical shadows}: The classical shadow prediction errors appear to be more narrowly distributed compared to the ML-based approaches; however, these predictions require measurements and do not include a learning step. In contrast, our approach encodes an entire family of states, and is capable of predicting arbitrary properties without necessitating further re-training or measurements. Nevertheless, we see that our approach performs comparably also to the shadow baseline.
\end{itemize} 

\begin{figure}
    \centering
    \subfloat[Second-order Rényi subsystem entanglement entropies]{\label{subfig:heisenberg-entropies}\includegraphics[width=\linewidth]{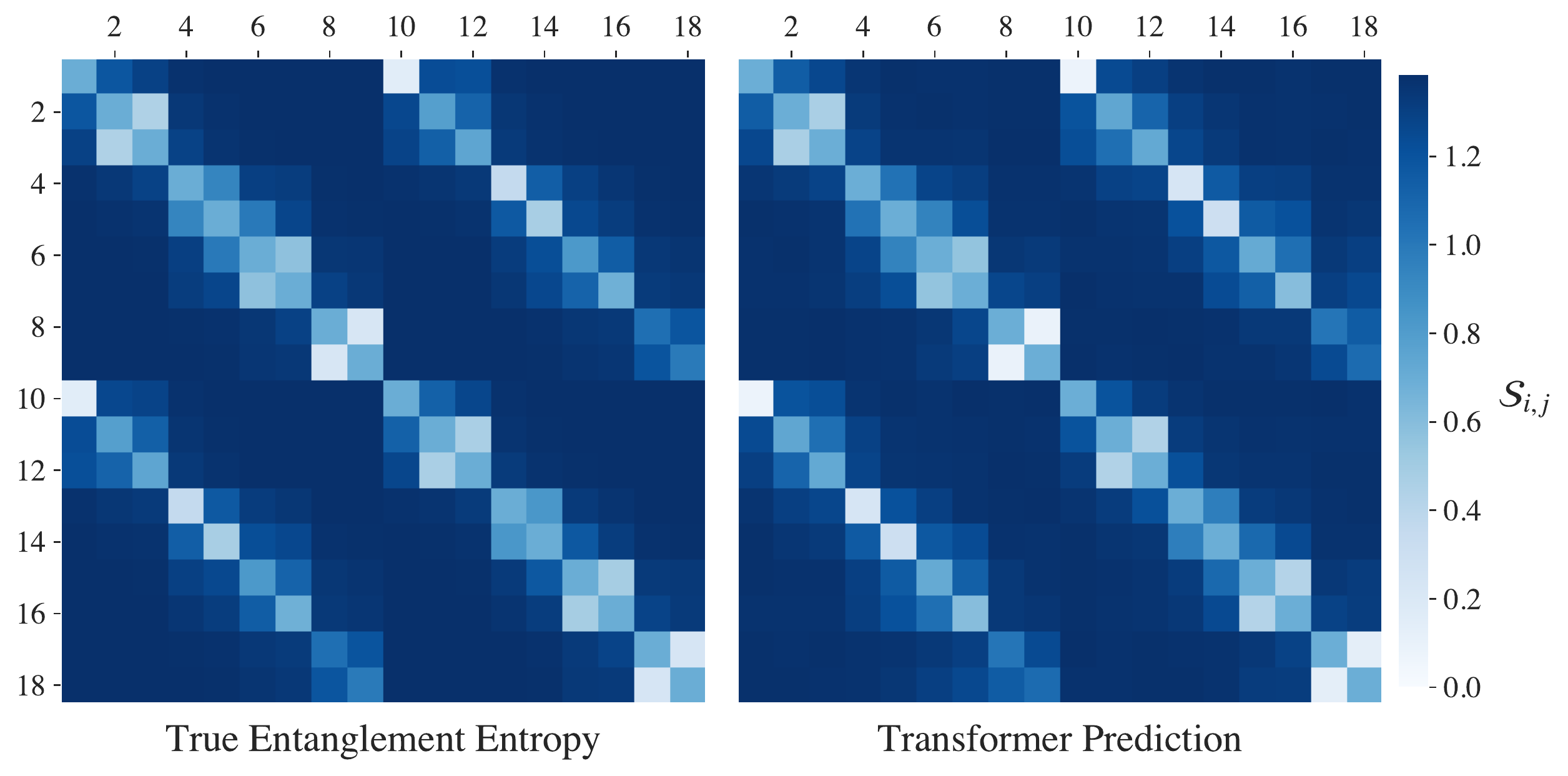}}\\
    \subfloat[Coupling graph]{\label{subfig:heisenberg-entropies-couplings}\includegraphics[width=\linewidth]{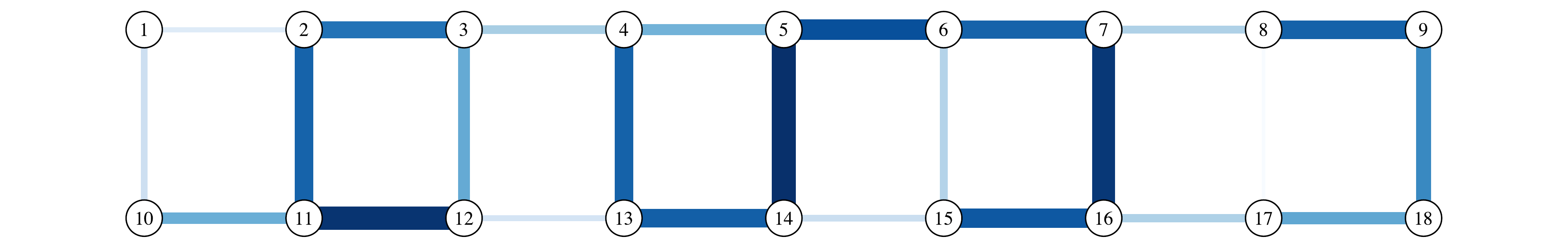}}
    \caption{a) Second-order Rényi subsystem entanglement entropies for subsystems of size at most two~(\ref{eq:renyi-entanglement-entropy}). The conditional generative model is conditioned on the coupling graph in (b), which determines the Hamiltonian~(\ref{eq:H-heisenberg}) of the 2D random Heisenberg model. The strength of the couplings is indicated by the width and color of the edges in the graph.}
    \label{fig:entanglement-entropy}
\end{figure}

\subsection{Predicting subsystem entanglement entropy}
\begin{figure}
    \centering 
    \includegraphics[width=.9\linewidth]{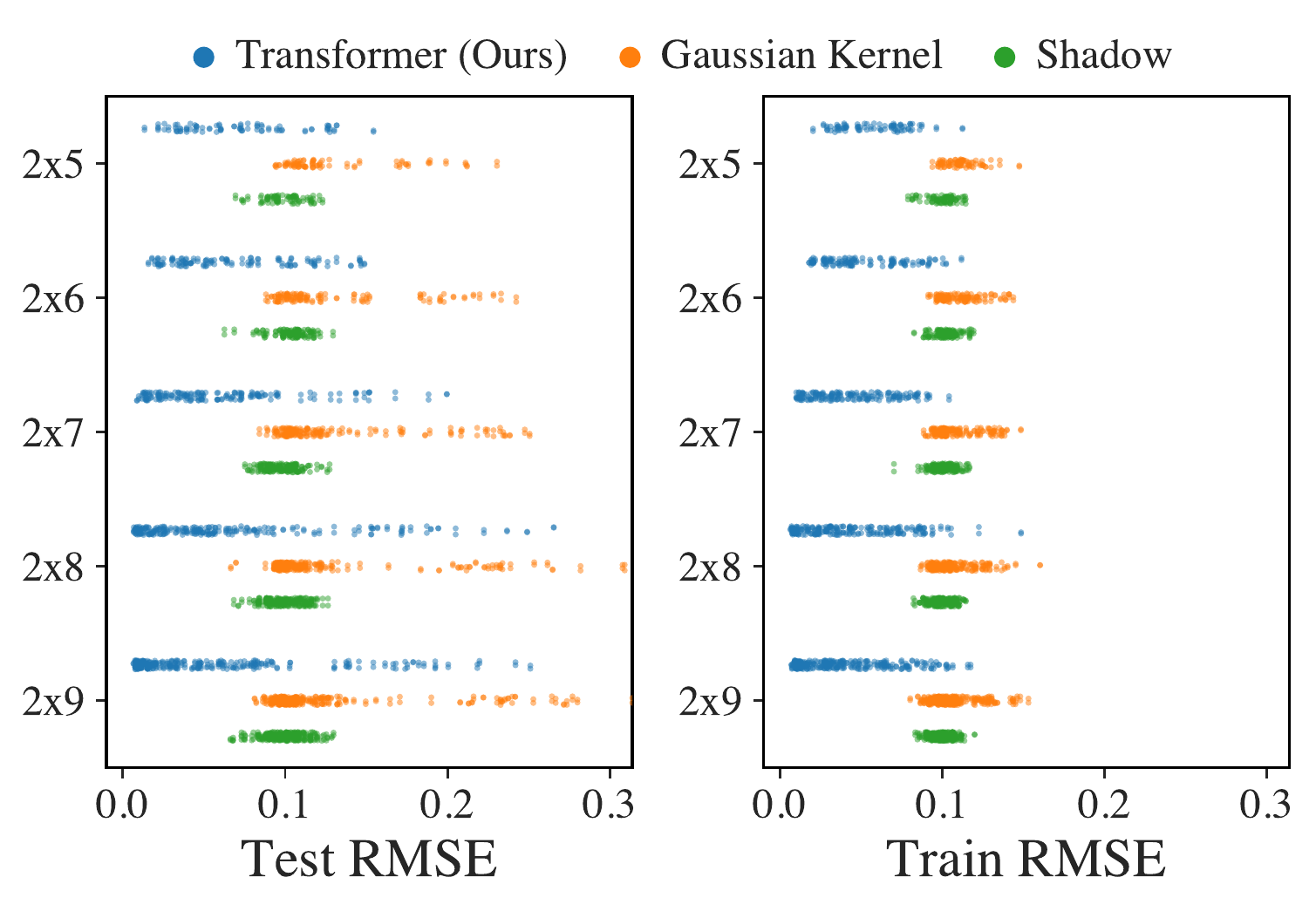}
    \caption{RMSE between true and predicted entanglement entropy for subsystems of size at most two, for our conditional generative model (blue), Gaussian kernel (orange), and shadow tomography (green). For a given system size, each point in the plot corresponds to the error of the entropy prediction for a particular subsystem, averaged over Hamiltonians from the test set.}
    \label{fig:rmse-entropy}
\end{figure}

We now consider the second-order Rényi subsystem entanglement entropies for subsystems of size at most two. This property is defined by
\begin{equation}
    \label{eq:renyi-entanglement-entropy}
    \cS(\rho_A) = -\log \tr{\rho_A^2},
\end{equation}
where $\rho_A$ is the reduced density matrix of the subsystem $A$. Since this quantity can be rewritten in terms of the expectation value of the local swap operator, the statistical guarantees of the classical shadow protocol imply that the required number of samples scales exponentially only in the subsystem size~\cite{huang2020predicting}, allowing one to compute $\cS(\rho_A)$ efficiently using classical shadows for small subsystems.
Fig.~\ref{fig:entanglement-entropy} shows the true and predicted entanglement entropies for the Hamiltonian defined by the (randomly sampled) coupling graph shown in Fig.~\ref{subfig:heisenberg-entropies-couplings}. The predictions are obtained analogously to the correlation functions (i.e., sampling from the conditional generative transformer, constructing the classical shadow, and predicting the property based on the shadow). Similar to the correlation functions, we see from the figure that the model makes fairly accurate predictions in comparison with the true entanglement entropies.
Fig.~\ref{fig:rmse-entropy} shows the distribution of RMSE for the entanglement entropies. Each point corresponds to a particular subsystem and shows the averaged RMSE over Hamiltonians from the test set (left panel) and the train set (right panel), respectively. Similar to the correlation functions, our approach leads to RMSEs with a higher variance compared to classical shadows, although here the distribution is skewed more towards zero, implying that our predictions are more accurate than the classical shadow predictions. Finally, in comparison with the Gaussian kernel, we can see that our generative model approach performs significantly better, throughout all system sizes.
Interestingly, these observations hold both for the training and the testing errors.

\section{Problem II: Rydberg atom systems}\label{sec:rydberg}

\begin{figure}[t]
\centering
    \includegraphics[width=\linewidth]{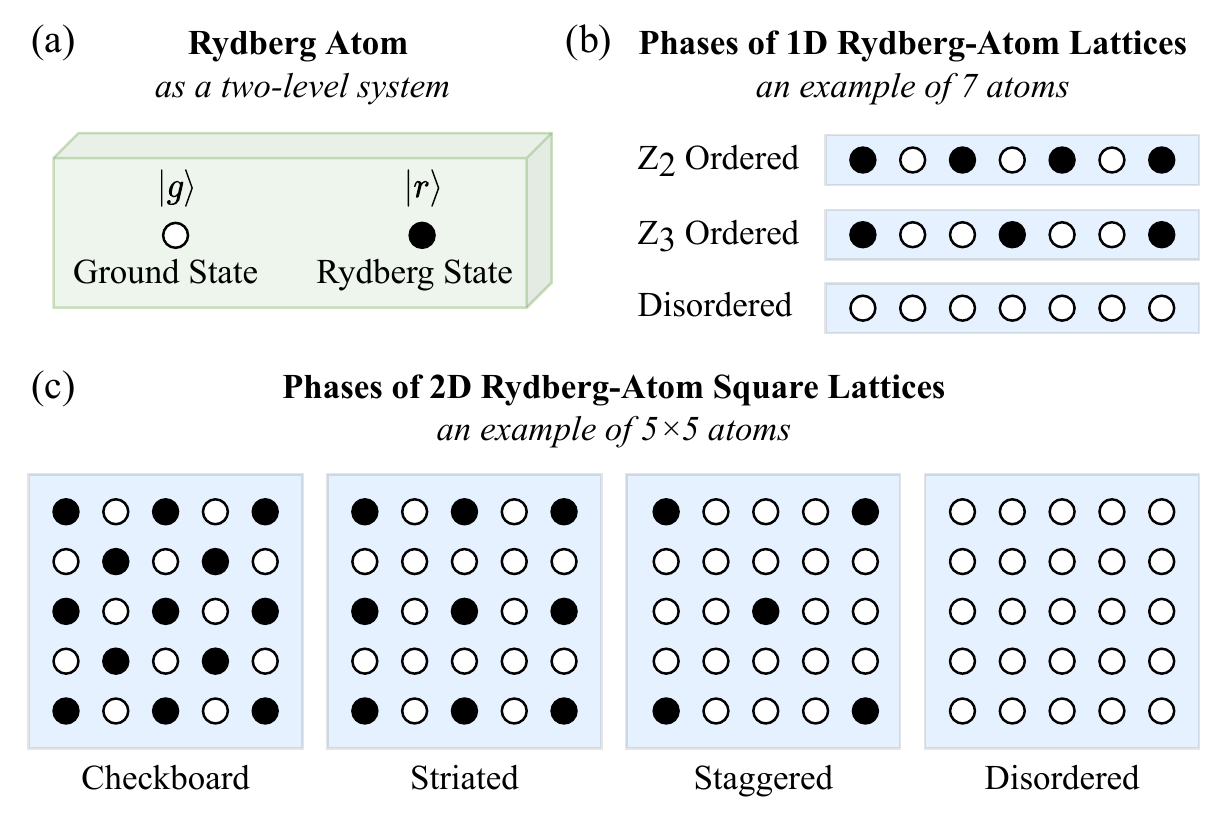}
    \vspace{-2em}
    \caption{Rydberg atom systems. (a) A Rydberg atom can be viewed as a two-level system, with the ground state $|g\rangle$ and Rydberg state $|r\rangle$ (an excited state). (b) Three quantum phases of 1D lattices of equidistant Rydberg atoms, in an example of 7 atoms. (c) Four quantum phases of 2D square lattices of Rydberg atoms, in an example of 5×5 atoms (the staggered phase shown here is a little different from that of Ref.~\cite{Rydberg2D-numerical} due to the boundary effect of this relatively small 5x5 lattice).}
    \label{fig:rydberg-demo}
\end{figure}

\begin{figure}[t!]
    \centering
    \includegraphics[width=\linewidth]{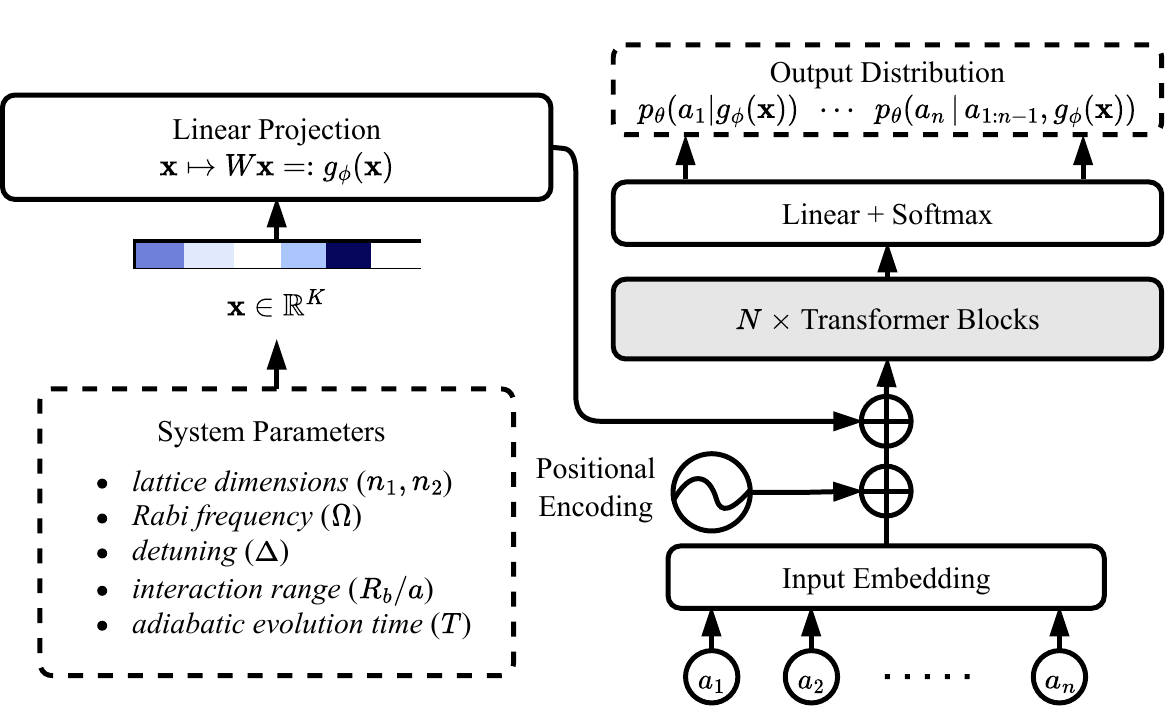}
    \caption{Structure of the conditional generative model for Rydberg atom systems.}
    \label{fig:rydberg-generative-model}
\end{figure}
The second family of Hamiltonians we consider is Rydberg atom systems. Trapped Rydberg atoms are a class of highly controllable neutral atoms that can be used to build programmable Ising-type quantum simulators~\cite{Rydberg-spin-liquids,Rydberg1D-51atoms,Rydberg2D-256atoms,scholl2021quantum,aquila,rydberg-logical}. In such simulations, a Rydberg atom is effectively a two-level system consisting of a ground state $\ket{g}$ and an excited state $\ket{r}$, where $\ket{r}$ is typically the $70S_{1/2}$ Rydberg state of neutral ${}^{87}$Rb atoms. Rydberg atoms can be allocated on a plane and trapped by light tweezers. In Ref.~\cite{Rydberg1D-51atoms}, Rydberg atoms are placed along a 1D lattice (i.e., a chain) equidistantly, while Ref.~\cite{Rydberg2D-256atoms} allocates the atoms on a 2D square lattice. For either a 1D or a 2D lattice of Rydberg atoms, the Hamiltonian of such a system can be written as
\begin{align}
\label{eq:H-rydberg}
    H=\frac{\Omega}{2} \sum_{i} X_{i}-\Delta \sum_{i} N_{i}+ \sum_{i<j}\frac{V_0}{|\vec x_i- \vec x_j|^6} N_{i} N_{j} ~,
\end{align}
where $\Omega$ is the Rabi frequency, $\Delta$ is the detuning of a laser, $V_0$ is the Rydberg interaction constant, and $\vec x_i$ is the position vector of the site $i$. $N_i = \ketbra{r_i}{r_i} = (1+Z_i)/2$ is the occupation number operator at site $i$, where $Z_i$ and $X_i=\ketbra{r_i}{g_i} + \ketbra{g_i}{r_i}$ are pseudo-Pauli operators. Following prior works \cite{Rydberg1D-51atoms,Rydberg2D-256atoms,huang2021provably}, we consider all atoms to be equidistantly distributed over a 1D chain or a 2D square lattice, and refer to the nearest-neighbor distance as the atom separation $a$. 
We define $R_0 = (V_0 / \Omega)^{1/6}$, and then use $R_0 / a$ as a parameter for the Rydberg-atom phase diagram (e.g., Fig. \ref{fig:real-rydberg-exp}) following the convention of prior studies \cite{Rydberg1D-51atoms,Rydberg2D-256atoms,huang2021provably}.

\textbf{Quantum Phases of Rydberg Atom Systems.}~The authors of Ref.~\cite{Rydberg1D-51atoms} experimentally show that for 1D chains of Rydberg atoms, the ground states can exhibit several different quantum phases, including the disordered phase, $Z_2$ ordered phase, and $Z_3$ ordered phase. On the other hand, it has been shown in Ref.~\cite{Rydberg2D-256atoms,Rydberg2D-numerical} that the ground states of 2D square lattices of Rydberg atoms also have several quantum phases, such as disordered, checkboard, striated, staggered and star phases. We provide illustrations for the cases of 1D and 2D in Fig. \ref{fig:rydberg-demo}.\looseness=-1

\textbf{Ground-State Preparation on a 256-Qubit Neutral-Atom Quantum Computer.}~We conduct quantum simulations of 13$\times$13 Rydberg atoms on Aquila \cite{aquila}, a 256-qubit neutral-atom quantum computer. Our goal is to reproduce the experiments of Ref.~\cite{Rydberg2D-256atoms}, which study the phase diagram of 2D Rydberg-atom systems through quantum simulations of 13$\times$13 Rydberg atoms. Following the experimental protocol of Ref.~\cite{Rydberg2D-256atoms}, we approximately prepare the ground state using adiabatic evolution \cite{albash2018adiabatic}.
Initially, each Rydberg atom is prepared in its own ground state $\ket{g}$, which globally corresponds to the ground state $\ket{g\ldots g}$ of the Hamiltonian \eqref{eq:H-rydberg} with a large negative detuning $\Delta$ and zero-valued $\Omega$. We then evolve the state under the time-dependent Hamiltonian \eqref{eq:H-rydberg} while linearly increasing the detuning strength $\Delta$, and subsequently the Rabi frequency $\Omega$, to desired positive values. The adiabatic theorem ~\cite{born1928beweis} implies that with a slow enough evolution, the final state should remain in the ground state of the final Hamiltonian. 
After preparing the ground state, we perform $Z$-basis measurement on all atoms simultaneously and collect the measurement readout results.

\textbf{Estimation Protocol.} 
Current experimental realizations of Rydberg quantum computers only support measurements in the computational basis (i.e., $Z$-basis), and therefore a hardware implementation would not be able to utilize the full classical shadow approach for state tomography. We instead follow the approach suggested by Ref.~\cite{torlai2019integrating}: without loss of generality, we may assume $\Omega > 0$ by a suitable global phase rotation, which (by the Perron-Frobenius theorem) implies the ground state can be chosen to have positive real coefficients. We will therefore make the approximation that the ground state preparation procedure produces a state with real amplitudes. This should be a good approximation if the adiabatically-prepared state is close to the ground state. This allows the states we study to be uniquely characterized by measurements in the computational basis.

\textbf{Model Structure.}~ In contrast to the 2D random Heisenberg models in Sec.~\ref{sec:heisenberg}, a Rydberg atom system can be described by scalar parameters such as $\Omega, \, \Delta$ and the lattice dimensions, which can be concatenated into a vector $\mathbf{x}$. For this reason, we directly use a linear model that takes the vector $\mathbf{x}$ as input and maps it to an embedding $g_\phi(\mathbf{x})$. Similar to the Heisenberg model, we then condition a transformer on $\mathbf{x}$ by adding $g_\phi(\mathbf{x})$ to the embedding right after the positional encoding. We illustrate this model structure in Fig. \ref{fig:rydberg-generative-model}.

\subsection{Predicting Quantum Phases of 2D Rydberg-Atom Square Systems (13$\times$13 atoms)}\label{sec:rydberg:aquila}

\begin{figure*}
    \centering
    \subfloat[Lattices]{\label{subfig:aquila-demo}\includegraphics[width=.15\linewidth]{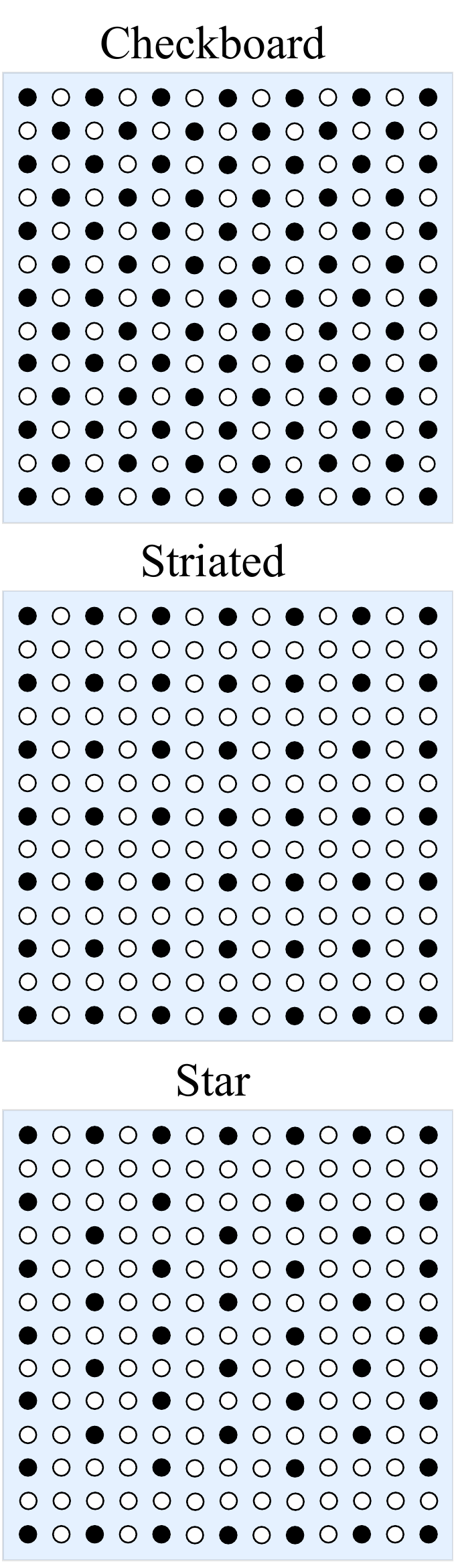}}%
    \hfill
    \subfloat[Training Samples]{\label{subfig:aquila-samples}\includegraphics[width=.25\linewidth]{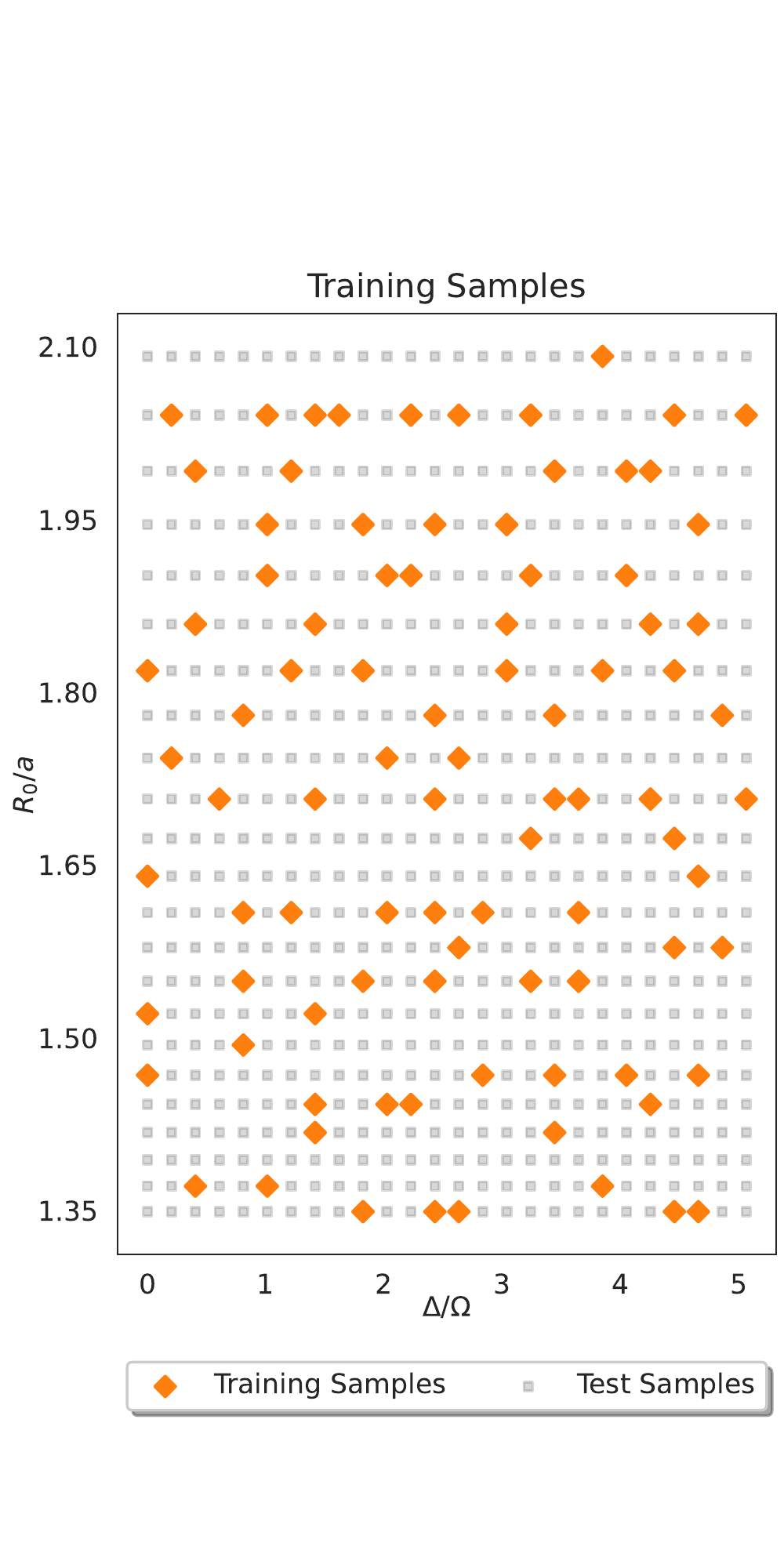}}%
    \hfill
    \subfloat[Phase Diagrams]{\label{subfig:aquila-phases}\includegraphics[width=.55\linewidth]{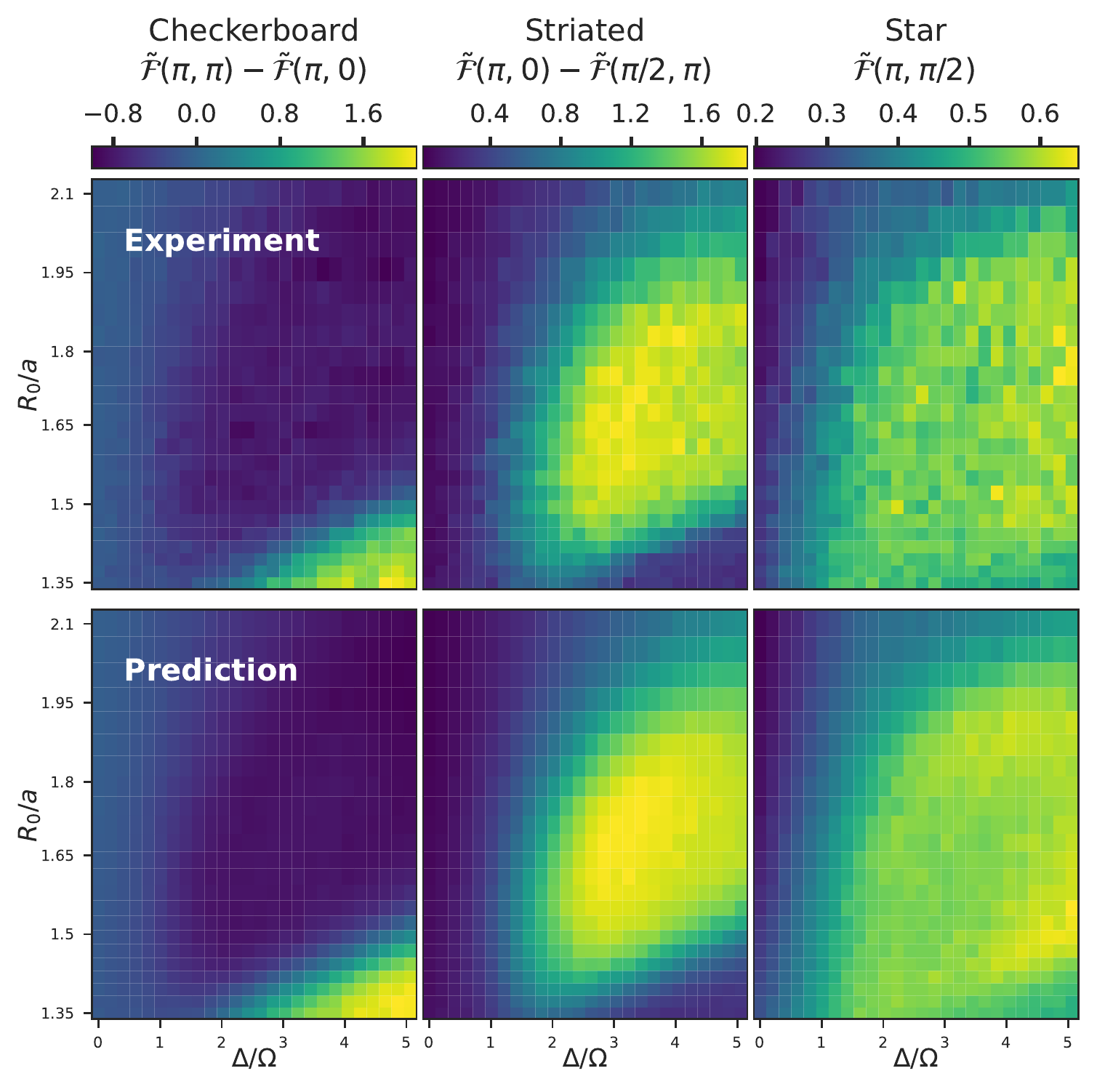}}%
    \caption{Phase Prediction of 13$\times$13 Rydberg-Atom Systems Prepared on Aquila \cite{aquila}, a neutral-atom quantum computer. \textbf{a)} Three phases under study. \textbf{b)} 19 training systems randomly sampled from 598 different configurations of Rydberg-atom systems. \textbf{c)} Phase diagrams computed with different order parameters on experimental data (upper row) and data generated by our model (lower row).
    }\label{fig:real-rydberg-exp}
\end{figure*}

Ref.~\cite{Rydberg2D-256atoms} studies three phases of the 13$\times$13 Rydberg-atom system: Checkboard, Striated, and Star phases. We apply the order parameters of these phases defined in Ref.~\cite{Rydberg2D-256atoms}, and obtain the phase diagram with our measurement data in Fig.\ref{subfig:aquila-phases}. The hardware constraints of Aquila \cite{aquila} are stricter than those of the device used in Ref.~\cite{Rydberg2D-256atoms}, preventing us from perfectly reproducing the phase diagram of Ref.~\cite{Rydberg2D-256atoms}. However, one can still clearly see that our phase diagram is quite close to the one obtained in Ref.~\cite{Rydberg2D-256atoms}.

\textbf{Dataset.}~The hardware constraint of Aquila only allows us to set up 13$\times$13 Rydberg atoms of nearest-neighbor spacing $a=4.0, 4.1, \dots, 6.2 \mu m$, and there are 23 spacing choices in total. For each $a$, we adjust the adiabatic evolution schedule to obtain $\Delta/\Omega = 0.1, 0.2, \dots, 1.0$ (26 equidistant numbers), respectively. In total, we have $23\times26=598$ 2D Rydberg atom square lattices of different $(a, \Delta/\Omega)$ configurations. For each system, we repeat the ground state preparation (via adiabatic evolution) 100 times and measure the prepared atoms in the computational basis simultaneously. We have $23\times26\times100=59,800$ measurement outcomes in total. Overall, we launched 598 quantum tasks with 100 shots per task on Amazon Braket\footnote{\url{https://aws.amazon.com/braket/}} \cite{braket}. Since the systems are all square lattices, which enjoy $C_4$ and reflection symmetries, we can apply data augmentation to the measurement data, leveraging the two symmetries, leading to 8x data size (478,400).

\textbf{Training.}~We randomly select 90 out of a total of 598 systems (15.1\%) to construct the training set, where each system comes with 800 measurement data points after data augmentation, summing to 72,000 training data points. The sampled systems for training are visualized in Fig. \ref{subfig:aquila-samples}. We adopt the same transformer structure as the one in Sec. \ref{sec:heisenberg}, and train it from scratch on the training data for 100,000 steps.

\textbf{Evaluation Results.}~We evaluate the trained model by letting it generate measurements for all 598, and apply order parameters defined in Ref.~\cite{Rydberg2D-256atoms} to obtain predicted phase diagrams for Checkboard, Striated, and Star phases in Fig. \ref{subfig:aquila-phases}. We also apply the order parameters to the real experimental measurement results we obtained from Aquila to obtain phase diagrams as the ground-truth in Fig. \ref{subfig:aquila-phases}. One can clearly see that the predicted phase diagrams are pretty close to the ground-truth. In addition, we provide quantitative evaluation results in Table \ref{tab:rydberg:phase-pred}, which measure the root-mean-square error (RMSE) for the predictions of different models relative to the ground-truth. Specifically, we compare our method against four baseline methods, including three kernel methods used in Ref.~\cite{huang2021provably} and a neural network method. Notice that the baseline methods train models on the order-parameter expectation values of the 90 training systems, while our model is trained directly on the measurements without knowledge of the order parameters. Clearly, our model significantly outperforms these baseline methods, corroborating the effectiveness of our proposed method. More details and results about these baseline methods are provided in Appendix \ref{apx:exp-details:rydberg:aquila}.

\begin{table}
\centering
\resizebox{.85\linewidth}{!}{
\begin{tabular}{l@{\hskip .2in}c@{\hskip .2in}}
    \toprule
    Method   &  Phase Pred. Error\\
\midrule
Dirichlet Kernel \cite{huang2021provably} & 0.2981 \\
    Gaussian Kernel \cite{huang2021provably}  & 0.0927 \\
    NTK \cite{huang2021provably}  & 0.0774  \\
    Neural Net  & 0.0647  \\
    Ours & \textbf{0.0569} \\
\bottomrule
\end{tabular}
}
\caption{
Comparison of Phase Prediction Errors. This table compares the Root-Mean-Square Error (RMSE) of order parameter value predictions for various models, including three kernel methods as used in Ref.~\cite{huang2021provably}, one neural network method, and our model. Our method achieves the lowest error rate.} 
\label{tab:rydberg:phase-pred}
\vspace{-1em}
\end{table}

\subsection{Proof-of-Concept Examples}\label{sec:rydberg:proof-of-concept}

In addition to the experiments on the Aquila quantum computer \citep{aquila}, we also run classical simulations of different Rydberg atom systems to conduct proof-of-concept examples demonstrating the power of our model.

\subsubsection{Predicting phases of larger quantum systems}\label{sec:rydberg:larger-systems}

Modern quantum devices can only simulate intermediate-scale quantum systems, while physicists are interested in systems of larger scales in many cases. Here, we use 1D Rydberg lattices as a showcase to demonstrate that our method can be used to predict properties of larger quantum systems (i.e., larger than any system in the training set). Specifically, we consider the phase prediction task for ground states of 1D Rydberg lattices. We collect ground-state measurements from 1D Rydberg lattices of $13,15,17,\dots,2k+1,\dots,29$ atoms, and train our model over these data. Then, we ask the model to predict the phase diagram of 1D Rydberg lattices of 31 and 33 atoms, which are larger than those in the training set.

\begin{figure*}[t!]
    \centering
    \subfloat[Ground-Truth]{\label{fig:larger-rydberg:true-phases} \includegraphics[width=\linewidth]{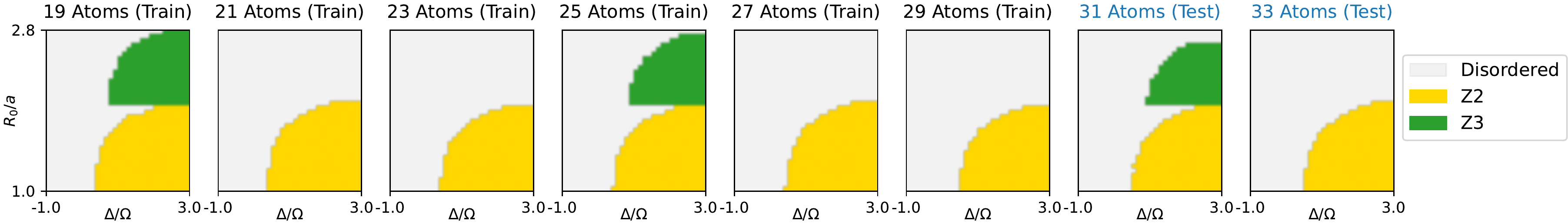}}
    \hfill
     \centering
    \subfloat[Prediction (Ours)]{\label{fig:larger-rydberg:ours-phases}\includegraphics[width=0.33\linewidth]{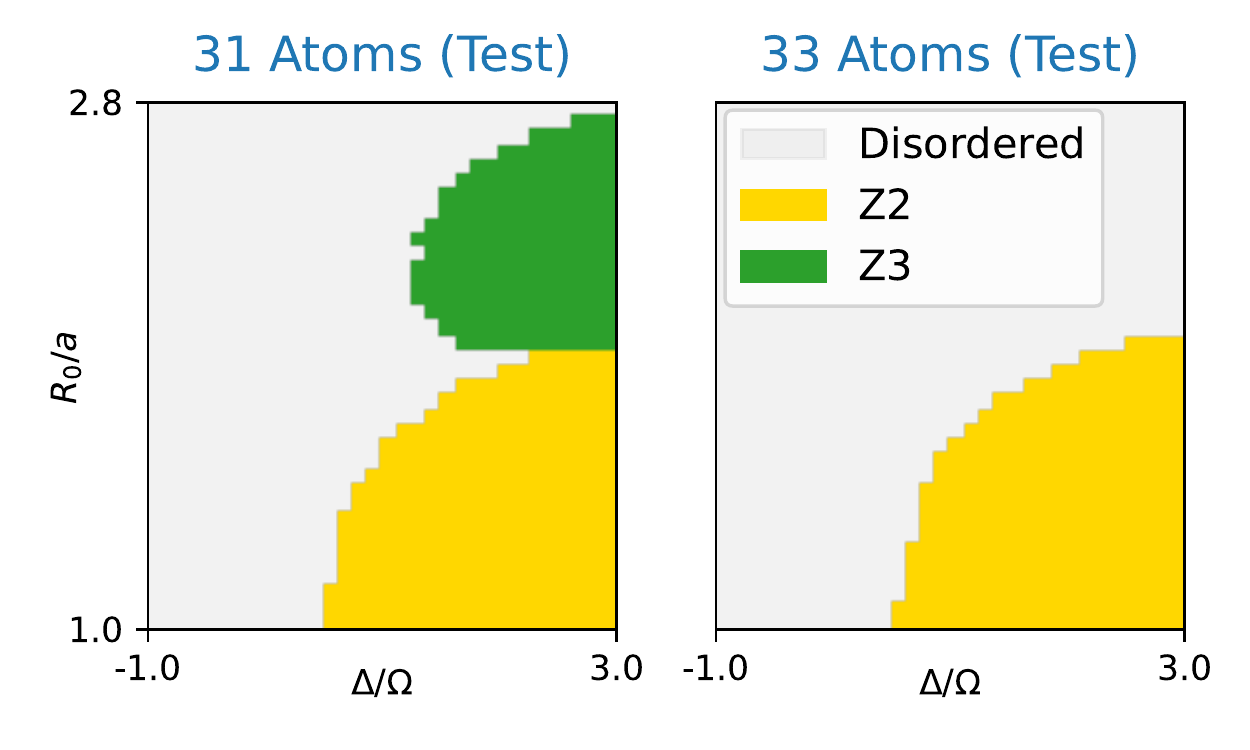}}
    \hfill
    \subfloat[Prediction (Gaussian Kernel)]{\label{fig:larger-rydberg:rbf-phases}\includegraphics[width=0.33\linewidth]{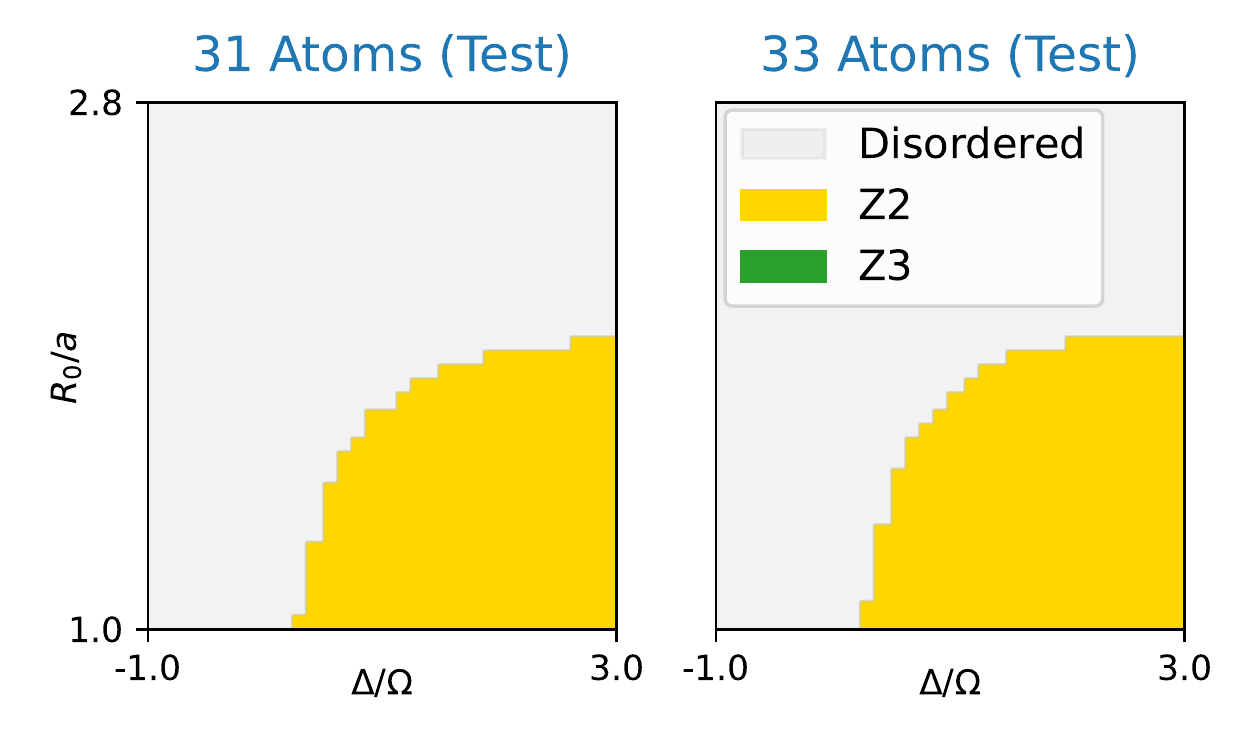}}
    \hfill
    \subfloat[Prediction (NTK)]{\label{fig:larger-rydberg:ntk-phases}\includegraphics[width=0.33\linewidth]{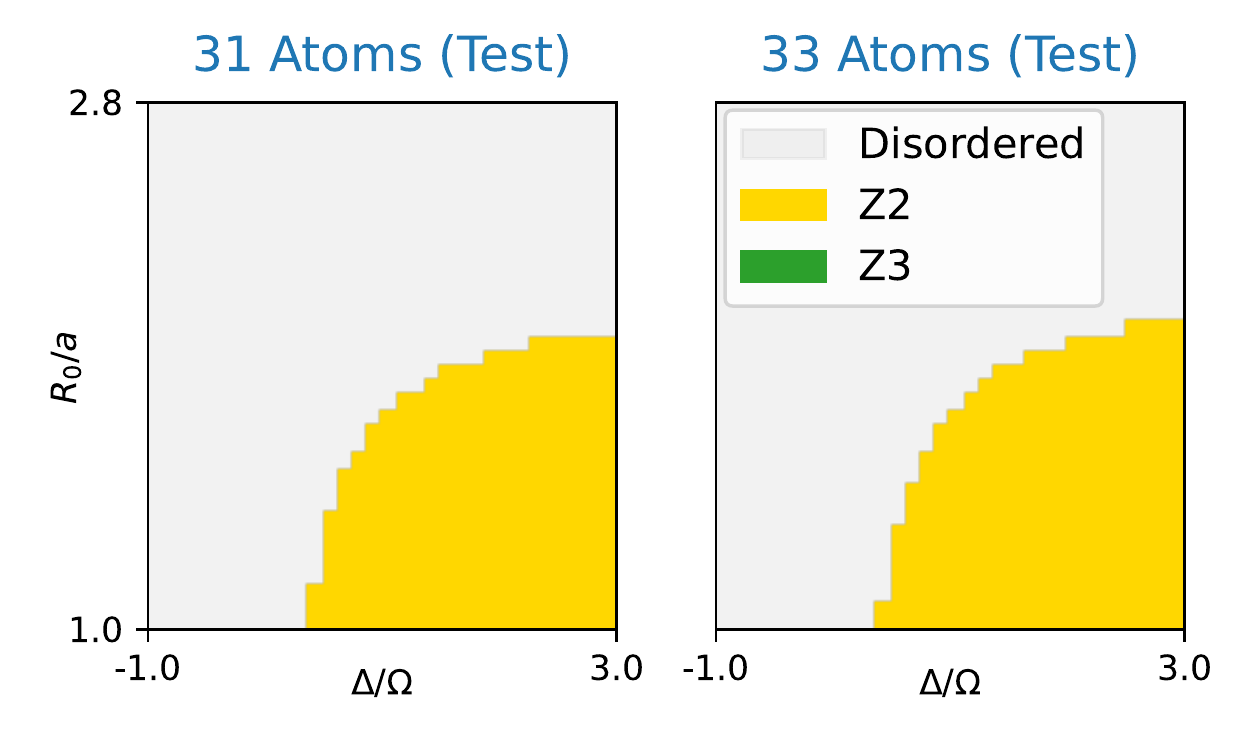}}
    \caption{Phase predictions for 1D Rydberg atom lattices of larger system sizes. (a) Ground-truth phase diagrams of 1D Rydberg atom lattices of increasing sizes, where the training set contains lattices of 13, 15, ..., 29 atoms \& the test set includes lattices of 31 and 33 atoms. (b) The predicted phase diagrams of our method. Note that our method accurately predicts the emergence and disappearance of the $Z_3$ phase in the 31-atom and 33-atom lattices, respectively. (c)(d)The predicted phase diagrams of two kernel methods used in \cite{huang2021provably}, which fail at predicting the emergence of the $Z_3$ phase for the 31-atom lattice.}
    \label{fig:rydberg-larger-systems} 
\end{figure*}

The true phase diagrams of these lattices are plotted in Fig. \ref{fig:larger-rydberg:true-phases}.
A special phenomenon one can easily observe from Fig. \ref{fig:larger-rydberg:true-phases} is the periodic emergence of the $Z_3$ phase; the $Z_2$ phase stably appears in the phase diagrams for lattices of $2k+1$ atoms, and the $Z_3$ phase only stably appears as the lattice has $3k+1$ atoms, where $k$ is any positive integer. Hence, to predict phases accurately for larger systems, a machine learning model must discover and understand this phenomenon.

\textbf{Dataset.}~To generate a dataset for training and evaluation, we use the \texttt{Bloqade.jl}~\cite{Bloqade} package to prepare ground states of Rydberg Hamiltonians with adiabatic evolution (i.e., tuning $\Omega$ and $\Delta$ gradually across time), following practices of prior experimental works \cite{Rydberg1D-51atoms,Rydberg2D-256atoms}.
Since ground states of Rydberg systems can be fully determined in the computational basis, for each numerically prepared state, we generate 1000 measurements in the computational basis. Overall, we conduct simulations for 1D and 2D Rydberg atom lattices, varying the lattice dimensions, atom separation $a$ (i.e., the distance between nearest neighbors), and the total adiabatic evolution time $T$. More details about the numerical simulations (e.g., schedulers of $\Omega$ and $\Delta$) are provided in Appendix \ref{apx:simulation:rydberg}.

From Fig. \ref{fig:larger-rydberg:ours-phases}, we can see that our model predicts (i) the emergence of the $Z_3$ phase for 31 atoms and (ii) the disappearance of the $Z_3$ phase for 33 atoms. Notice that our model only saw $Z_3$ phases only for 19 and 25 atoms during training.
This result implies that our proposed model is indeed able to predict certain properties for systems whose sizes exceed the training systems. In comparison, Fig. \ref{fig:larger-rydberg:rbf-phases} and \ref{fig:larger-rydberg:ntk-phases} show that the kernel methods used in Ref.~\cite{huang2021provably} cannot predict the emergence of the $Z_3$ phase for 31 atoms, after being trained on the same dataset as our model.

\begin{figure*}[t!]
    \centering
    \subfloat[Ground-Truth]{\label{fig:2d-longer-time-phase:true}\includegraphics[width=.195\linewidth]{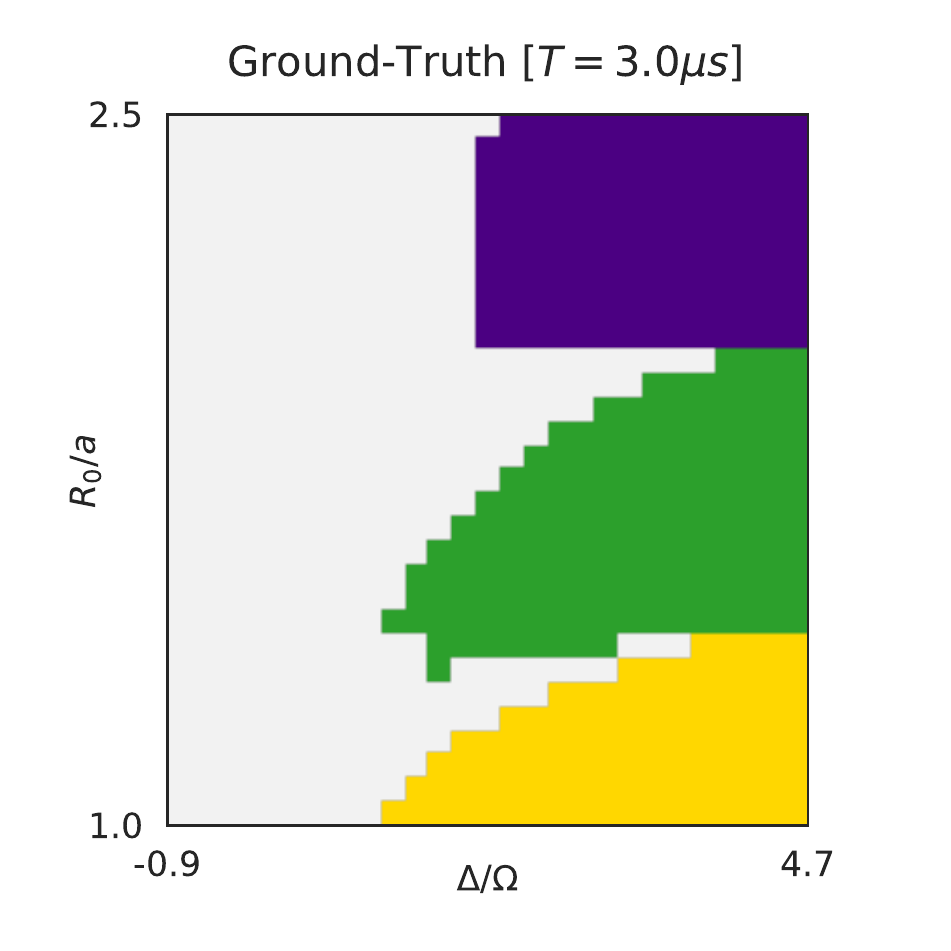}}%
    \hfill
    \subfloat[Prediction (Ours)]{\label{fig:2d-longer-time-phase:ours}\includegraphics[width=.195\linewidth]{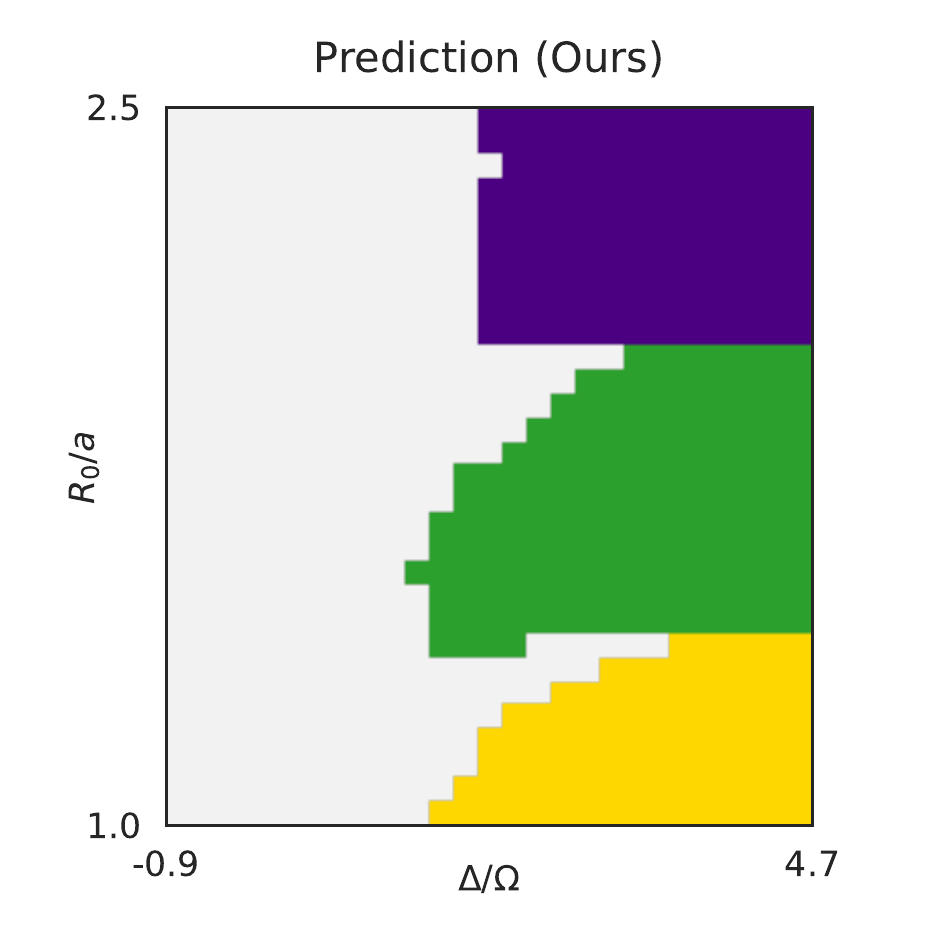}}%
    \hfill
    \subfloat[Prediction (Baseline) ]{\label{fig:2d-longer-time-phase:baseline}\includegraphics[width=.195\linewidth]{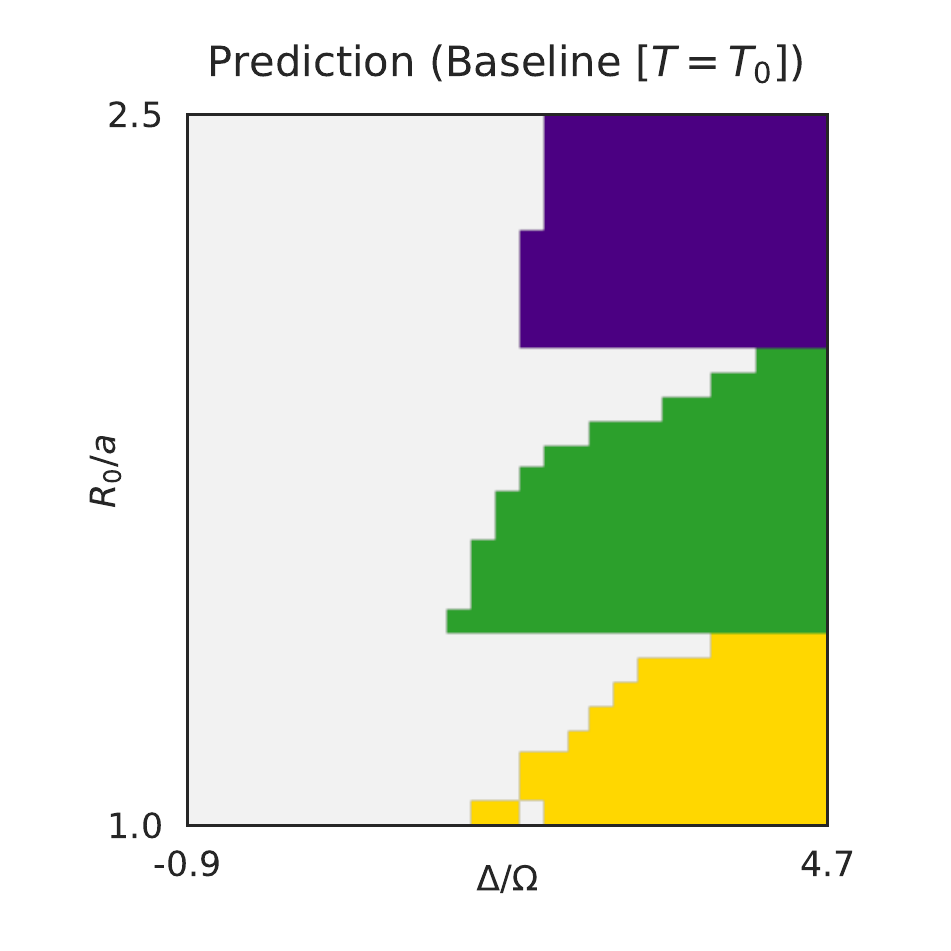}}%
    \hfill
    \subfloat[Prediction (Gaussian Kernel) ]{\label{fig:2d-longer-time-phase:rbf}\includegraphics[width=.195\linewidth]{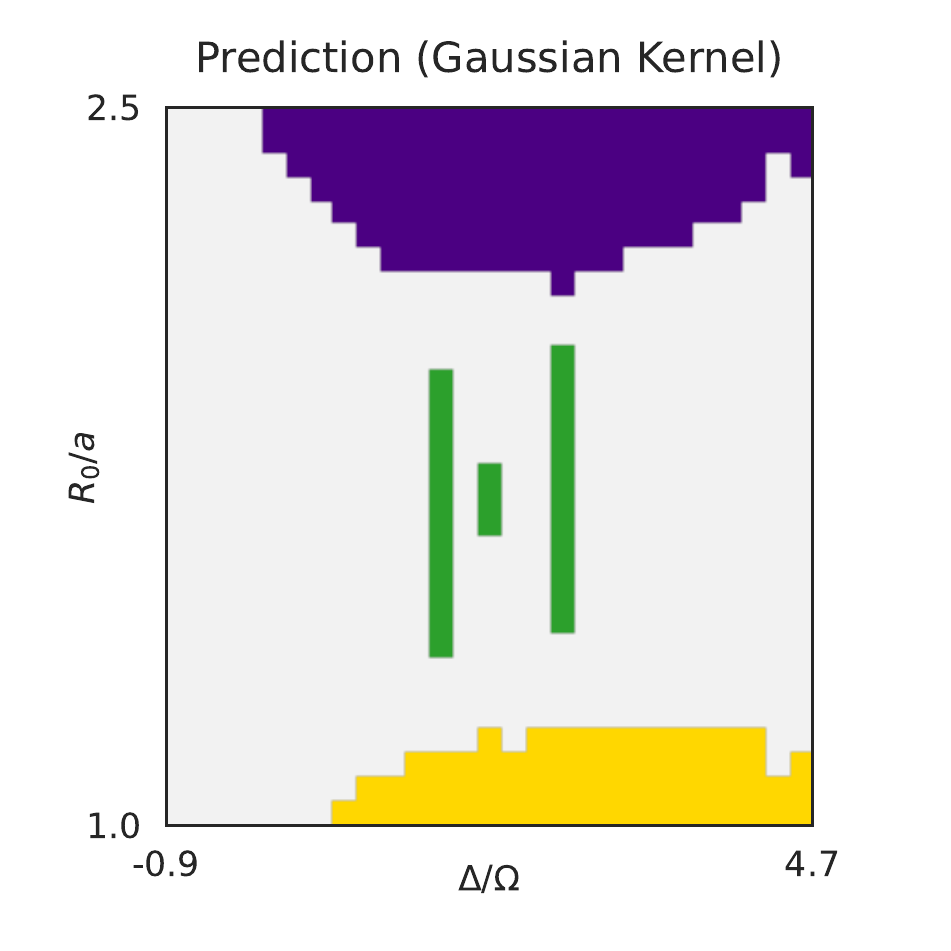}}%
    \hfill
    \subfloat[Prediction (NTK)]{\label{fig:2d-longer-time-phase:ntk}\includegraphics[width=.195\linewidth]{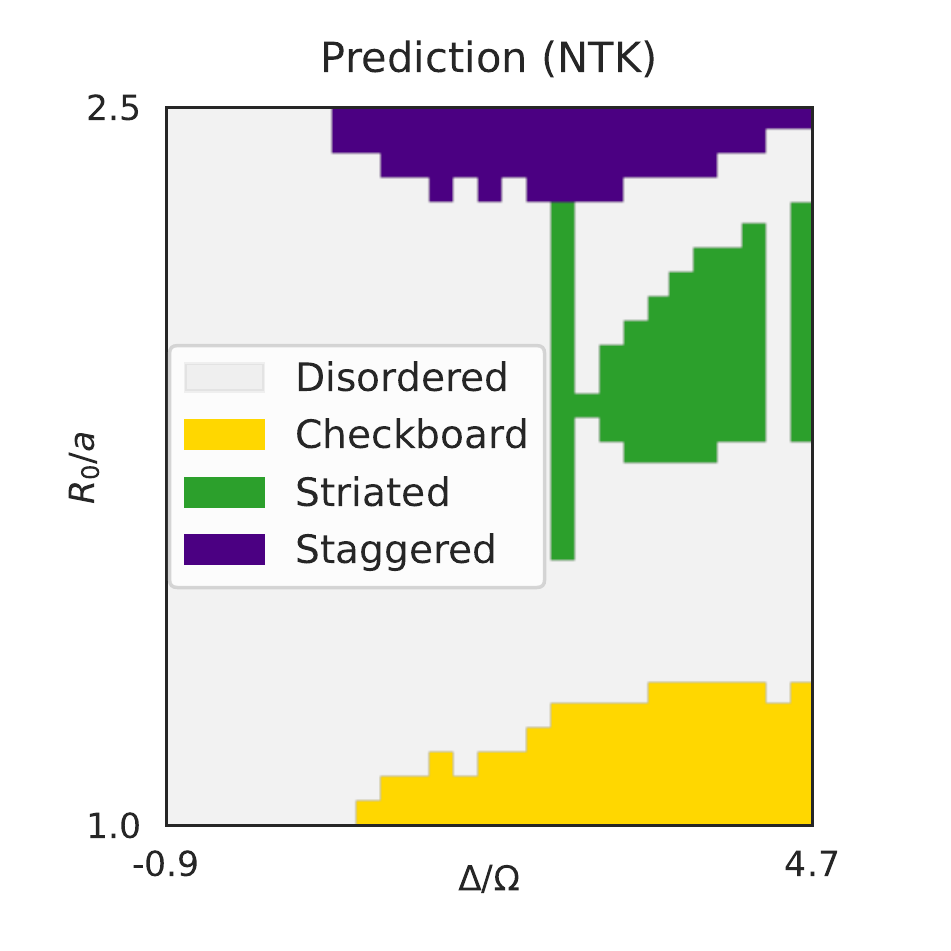}}%
    \caption{Phase predictions for the 2D Rydberg atom lattice (5x5 atoms) as an adiabatic evolution of $T=3.0 \mu s$, while the models are being trained over lattices with adiabatic evolution time $T\leq T_0 =1.0 \mu s$. a) Ground-truth phase diagram as $T=3.0 \mu s$. b) Predicted phase diagram by our conditional generative model. c) Predicted phase diagram by the \texttt{Baseline [$T=T_0$]} method, which uses the ground-truth at $T=T_0$ as the prediction for larger $T$. d) Predicted phase diagram by Gaussian kernel. e) Predicted phase diagram by Neural Tangent Kernel (NTK).}\label{fig:2d-longer-time-phases}
    \vspace{-1em}
\end{figure*}

\begin{figure}[t!]
    \centering
    \includegraphics[width=1\linewidth]{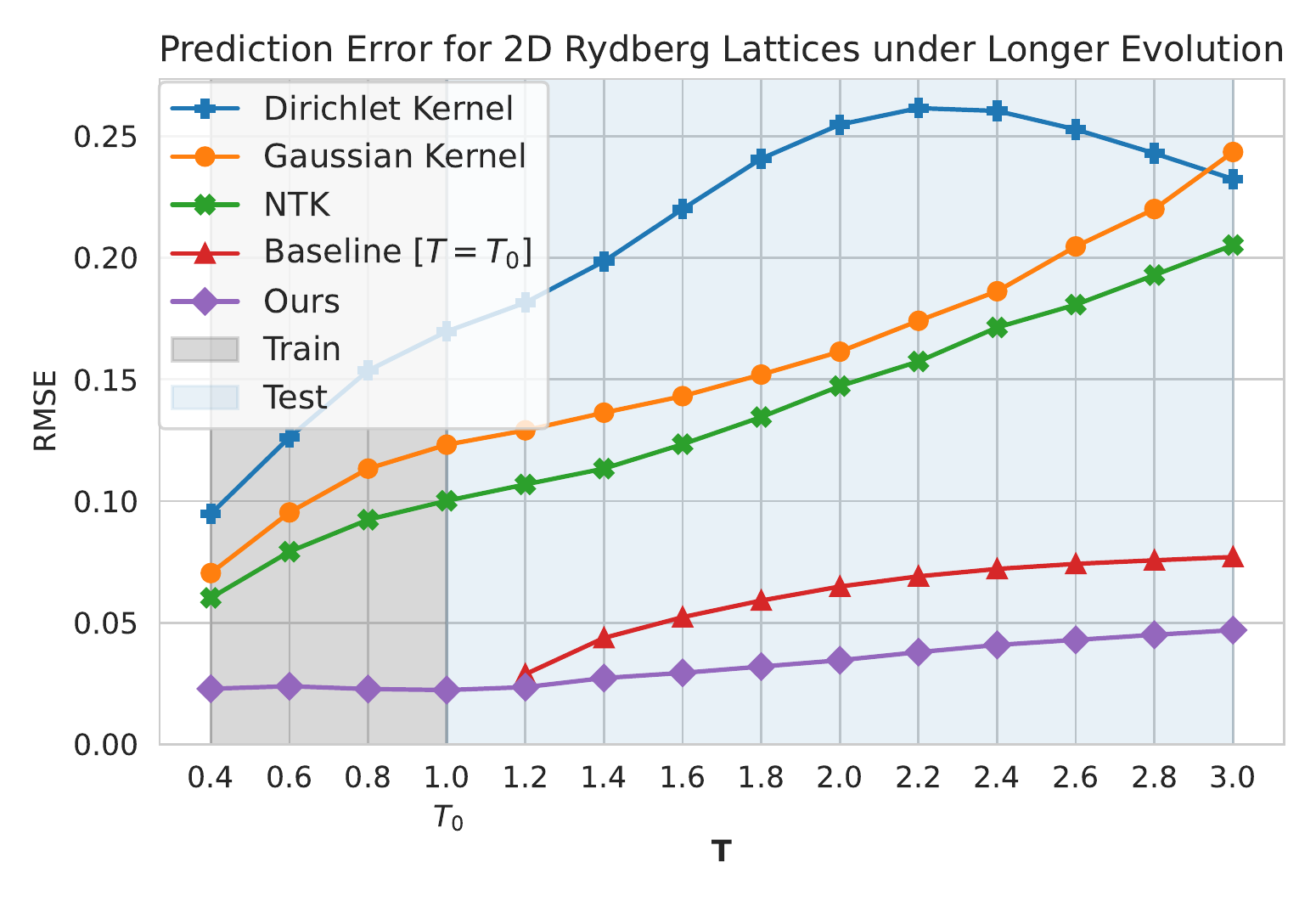}
    \vspace{-1em}
    \caption{Average prediction errors of phase order parameters for a 2D Rydberg lattice (5x5 atoms) simulated with longer adiabatic evolution times than $T_0 = 1.0 \mu s$. Three order parameters, $O_{\mathrm{checkboard}},O_{\mathrm{striated}},O_{\mathrm{staggered}}$ (see Appendix \ref{apx:exp-details:rydberg:order-params} for definitions), are predicted, and the prediction errors are computed in RSME. Our method is compared with a baseline method, \texttt{Baseline [$T=T_0$]} (using the ground-truth at $T=T_0$ as the prediction for larger $T$), and three kernel methods used in Ref.~\cite{huang2021provably}, Dirichlet, Gaussian, and neural tangent kernel (NTK), are compared in this experiment. 
    }
    \label{fig:rydberg-longer-systems}
    \vspace{-1em}
\end{figure}

\subsubsection{Predicting phases of ground states prepared with longer adiabatic evolution time}\label{sec:rydberg:longer-evolution}

The ground-state preparation of Rydberg atoms via adiabatic evolution needs a sufficiently long evolution time to reach the exact many-body ground state. Otherwise, the prepared state becomes an excited state (possibly of low energy) instead. Typically, the longer the evolution is, the more accurate the preparation becomes. However, currently, Rydberg-atom experiments can only endure for a few microseconds \cite{quera-coherent-transport,quera-MIS}, leading to inaccurate adiabatically prepared states in many cases. Thus, predicting the properties of states prepared with longer adiabatic evolution times can give practitioners a more accurate understanding of the target ground state.

Here, we consider a 2D Rydberg square lattice of 5x5 atoms, and study the predictive performance of our model for states prepared with longer evolution times.
The training set contains measurements for systems simulated with the total adiabatic evolution time, $T \leq T_0 := 1.0\,\mu s$. 
The trained conditional generative model is then used to predict phases of systems simulated with evolution times $T > T_0 = 1.0\,\mu s$. 
In Fig.~\ref{fig:rydberg-longer-systems}, we plot the average prediction error of the trained model for the phase order parameters, and compare it with the baseline that keeps using the ground-truth phase diagram at $T=T_0$\footnote{This is the practice in real Rydberg atom experiments \cite{Rydberg2D-256atoms}.}. We can see from Fig. \ref{fig:rydberg-longer-systems} that our trained model is able to predict the phases of Rydberg atom systems simulated with longer adiabatic evolution times more accurately than this baseline method (named as \texttt{Baseline [$T=T_0$]} in Fig.~\ref{fig:rydberg-longer-systems}), which directly uses the ground-truth phase diagram at $T=T_0$ as prediction for any $T>T_0$. Also, our method outperforms kernel methods used in Ref.~\cite{huang2021provably} significantly. The predicted phase diagrams by our method and the kernel methods for $T=3.0 \mu s$ are provided in Fig.~\ref{fig:2d-longer-time-phases}, and one can easily observe that our method's prediction is much more accurate compared with the kernel methods used by Ref.~\cite{huang2021provably}.

\section{Discussion}
The results presented in this work provide tangible evidence that state-of-the-art classical machine learning models have the capacity to accurately represent entire families of quantum states. By using trainable embeddings to condition generative models on the classical parameters of the quantum system of interest, we have shown that these generative models can generate artificial measurement samples. These samples correspond to states that were not included in the training set and can be used to predict properties such as local observables, entanglement entropies, and phase diagrams. Our method has been demonstrated on the ground states of two-dimensional anti-ferromagnetic Heisenberg models and Rydberg atom systems, showing remarkable improvements over related techniques based on shadow tomography and kernel methods.
Throughout our studies, we have utilized the transformer architecture as our generative model due to its proven success in sequence modeling. However, the attention mechanism of transformers is an operation which scales quadratically with the length of the input sequence, which can become problematic when modeling long sequences of thousands of qubits. This presents exciting future research avenues where one could explore other models, such as long-range transformers \cite{peng2021random}, or other architectures designed to handle long-range dependencies, like state-space models \cite{gu2022efficiently,li2022makes}.


\paragraph*{Note.} During the completion of this work, a related approach has been released on arXiv~\cite{zhang2022transformer}, which, similar to our technique, uses transformers to represent a family of quantum states. Different from ours, they treat the conditioning variable as a separate element of the sequence and only consider vector-valued classical variables. Furthermore, they directly parametrize the wave function and minimize the variational energy of a Hamiltonian of interest. In this way, only pure states can be modeled, posing additional challenges when considering realistic quantum systems where noise and errors are prevalent.

\vspace{-1em}
\section*{Acknowledgments}
The authors thank Mao Lin, Peter Komar, Matthew Beach, Yiheng Duan, Xiuzhe (Roger) Luo, Hsin-Yuan Huang, Giacomo Torlai, Shahnawaz Ahmed, Nathan Killoran, Korbinian Kottmann, Di Luo, Hong-Ye Hu for valuable comments and inspiring discussions.

\newpage
\bibliography{refs-main}

\newpage
\onecolumngrid
\newpage
\appendix
\section{Model Structures}

\subsection{Transformer}
\label{apx:transformer-architecture}
Transformer networks constitute a powerful class of sequence models that use attention mechanisms as their central building blocks, and this structure has also been applied in various sub-fields of physics \cite{cha2021attention,carrasquilla2021probabilistic,luo2022autoregressive,wang2021spacetime}. In this work, we closely follow the original proposal of transformer networks \cite{vaswani2017attention}, adopting the standard encoder-only transformer architecture. The network consists of an embedding layer, followed by a sequence of $N$ decoder blocks with self-attention modules.

\paragraph{Embedding and Positional Encoding.} The transformer architecture uses learned embeddings that linearly map one-hot encoded input tokens to vectors of dimension $d_\mathrm{model}$. After the embedding, positional information is injected via a positional encoding layer. This mechanism allows the model to make use of the relative and absolute positions of tokens in a sequence. The positional encodings are given by sine and cosine functions of different frequencies and are functions of the dimension and position $k$ in the sequence
\begin{equation}
    pe(k,\,2i) = \sin\left(\frac{k}{10000^{2i / d_{\mathrm{model}}}}\right),
    \hspace{2em}
    pe(k,\,2i + 1) = \cos\left(\frac{k}{10000^{2i / d_{\mathrm{model}}}}\right),
\end{equation}
where $i$ refers to the dimension. After embedding and positional encoding, the first component of a decoder block is the self-attention mechanism. 

\paragraph{Multi-head Self-Attention.} An attention function can be described as mapping a query $Q$ and a set of key-value pairs $K,\,V$ to an output which is computed as a weighted sum with weights based on the query and key. The specific attention function used in the transformer architecture is called scaled dot-product attention and is given by the function
\begin{equation}
    \mathrm{Attention}(Q,\,K,\,V) = \mathrm{softmax}\left(\frac{QK^T}{\sqrt{d_k}}\right)V,
\end{equation}
where $Q,\,K$ and $V$ are linear transformations of the input vectors
\begin{equation}
    Q = XW_Q,\hspace{1em}
    K = XW_K,\hspace{1em}
    V = XW_V
\end{equation}
where $X\in\R^{n\times d_\mathrm{model}}$ is the matrix of the $n$ embedded input tokens with dimension $d_\mathrm{model}$. As in Ref.~\cite{vaswani2017attention}, we use multi-head-attention where the input vectors are linearly projected $n_h$ times to query, key, and value vectors resulting in $n_h$ attention vectors. These are then concatenated and again projected so that the final output of the multi-head self-attention module is
\begin{equation}
    \begin{aligned}
        \mathrm{Multi-head}(Q,\,K,\,V) &= \mathrm{Concat}(\mathrm{head}_1,\,\ldots,\,\mathrm{head}_{n_h})\\
        \mathrm{where} \, \mathrm{head}_i &= \mathrm{Attention}(Q_i,\,K_i,\,V_i)
    \end{aligned}
\end{equation}
with $Q_i = XW^{(i)}_Q$, $K_i = XW^{(i)}_K$ and $V_i = XW^{(i)}_V$.

\paragraph{Position-wise Feedforward.}
After the multi-head attention layer, we have a position-wise feedforward network, which is a fully connected neural network with two linear transformations and a ReLU activation in between, applied to each position separately and identically.\\

Besides, each sublayer (i.e., self-attention or position-wise feedforward) has a residual connection and is followed by layer normalization, resulting in the transform \texttt{LayerNorm($x$ + SubLayer(x))}.
Finally, after the last transformer block, the output is passed through a linear projection and a softmax layer.

\subsection{Conditioning Network}

In our proposed model structure, there is a conditioning network that takes the classical description of a quantum system and outputs a representation of this system. As we show in Fig. \ref{fig:heisenberg-generative-model} and \ref{fig:rydberg-generative-model}, one may choose different network architectures suitable for various types of quantum systems. Here we explain our two choices of conditioning network structures, graph neural networks (GNN) and linear models, for the problems of 2D random Heisenberg models and Rydberg atom systems, respectively.

\subsubsection{Graph Neural Network (GNN)}

In this paper, we adopt one of the most popular GNN structures, Graph Convolution Network (GCN) \cite{kipf2017semisupervised}, as the conditioning network for 2D random Heisenberg models.
Given a weighted undirected graph $\mathcal{G}$ with $n$ nodes and the adjacency matrix $A$, the GCN consists of $L$ graph convolutional layers. The $(l+1)$-th layer is given by
\begin{align}
H^{(l+1)}= \mathrm{ReLU} \left(\tilde{D}^{-\frac{1}{2}} \tilde{A} \tilde{D}^{-\frac{1}{2}} H^{(l)} W^{(l)}\right)
\end{align}
where $\tilde{A}=A+I_N$ ($I_N$ is the identity matrix), $\tilde D$ is a diagonal matrix with $\tilde{D}_{i i}=\sum_j \tilde{A}_{i j}$, $W^{(l)}$ is a layer-specific trainable weight matrix, $\mathrm{ReLU}(\cdot)=\max (0, \cdot)$ is an activation function and $H^{(l)} \in \mathbb{R}^{N \times D}$ is the output of the $l$-th layer. In our implementation, $H^{(0)}=X$, where $X\in \mathbb{R}^{N \times 1}$ with the $i$-th element as the weighted degree of the $i$-th node. After passing the graph $\mathcal G$ through the graph convolutional layers, we obtain node embeddings $z_1,\dots, z_n$, i.e., the rows of $H^{L}$). Subsequently, we apply a linear projection, with trainable weights $\{W_i\}_{i\in [n]}$, on the embeddings to obtain an encoding that will be fed to the transformer, 
\begin{align}
    z_1,\dots,z_n \mapsto \sum_{i=1}^n W_i z_i.
\end{align}

\subsubsection{Linear Model}
For Rydberg atom systems, since the classical description of a system is just a set of scalars, we find that a linear model suffices for the conditioning network. For instance, for a Rydberg atom system with $K$ system parameters, we concatenate the parameters into a vector $\mathbf{x}\in \mathbb R^K$, and apply the linear transformation
\begin{align}
    \mathbf{x} \mapsto W \mathbf{x} + b
\end{align}
where $W$ is a trainable weight matrix and $b$ is a trainable bias term.

\section{Experimental details}
\subsection{2D anti-ferromagnetic random Heisenberg model}
\label{apx:exp-details:heisenberg}
For 2D random Heisenberg models, our transformer-based generative model is trained for 100 epochs with mini-batches of size 100, using the Adam Optimizer \cite{adam}. We varied the learning rate using 5 warmup epochs, increasing the learning rate from 0 to \texttt{1e-3}, followed by a cosine decay with a maximum learning rate set to \texttt{1e-3} and final learning rate set to \texttt{1e-7}. We use dropout regularization with a dropout rate 0.1 \cite{dropout}. In the transformer architecture, we use 4 decoder blocks, with 4 attention heads and model dimensionality set to 128.
The graph neural network consists of 3 graph convolutional layers with ReLU activations and with hidden sizes 64, 32, and 16, respectively. After the GCN layers, a linear transformation is applied to each position separately, with dimension 128 (equal to the transformer dimension). All models are trained for approximately 2.5 hours on a single NVIDIA TITAN Xp GPU with 12GB of memory.

\subsection{Rydberg atom systems} \label{apx:exp-details:rydberg}

\subsubsection{Hyper-parameters}

For Rydberg atom systems, our transformer-based generative model is trained for 20k-50k iterations with mini-batches of size 512, using the Adam Optimizer. We initialize the learning rate with \texttt{1e-3}, followed by a cosine decay with the final learning rate set to \texttt{1e-7}. We use dropout regularization with a dropout rate 0.1. In the transformer, we use 4 decoder blocks with 4 attention heads and model dimensionality set to 128.
Each model is trained on a single GPU, which is either an NVIDIA V100 or NVIDIA A6000. 10k iterations take about 1 hour on a single GPU.

\subsubsection{More Details about Sec. \ref{sec:rydberg:aquila}}\label{apx:exp-details:rydberg:aquila}

\paragraph*{\textbf{2D Square Lattices.}}
Following Ref.~\cite{Rydberg2D-256atoms}, we define the order parameters for 2D square lattices of Rydberg atoms using the Fourier transformation. Specifically, we apply the Fourier transformation to single-shot measurement outcomes as
\begin{align}
    \mathcal{F}(\mathbf{k})=\Bigl| \sum_i \exp \left(i \mathbf{k} \cdot \vec x_i / a\right) N_i / \sqrt{n}\Bigr|
\end{align}
where $n$ is the number of atoms. Then, we use the symmetric Fourier transform 
\begin{align}
    \tilde {\mathcal F}(k_1,k_2) =  {\mathcal F}(k_1,k_2) + {\mathcal F}(k_2,k_1) / 2 
\end{align}
to take the reflection symmetry into consideration.

Following Ref.~\cite{Rydberg2D-256atoms} defines the order parameters for the Checkboard and Striated phases as 
\begin{align}
    O_{\mathrm{checkboard}} &= (\cF(\pi,\pi) - \tilde \cF(\pi,0)  ) \label{eq:order-param:checkboard}\\
    O_{\mathrm{striated}} &= (\tilde \cF(\pi,0) - \tilde \cF(\pi/2,\pi) )\label{eq:order-param:striated}\\
    O_{\mathrm{star}} &= \tilde \cF(\pi,\pi/2) \label{eq:order-param:star}
\end{align}

\begin{figure}[t!]
    \centering
    \includegraphics[width=.6\linewidth]{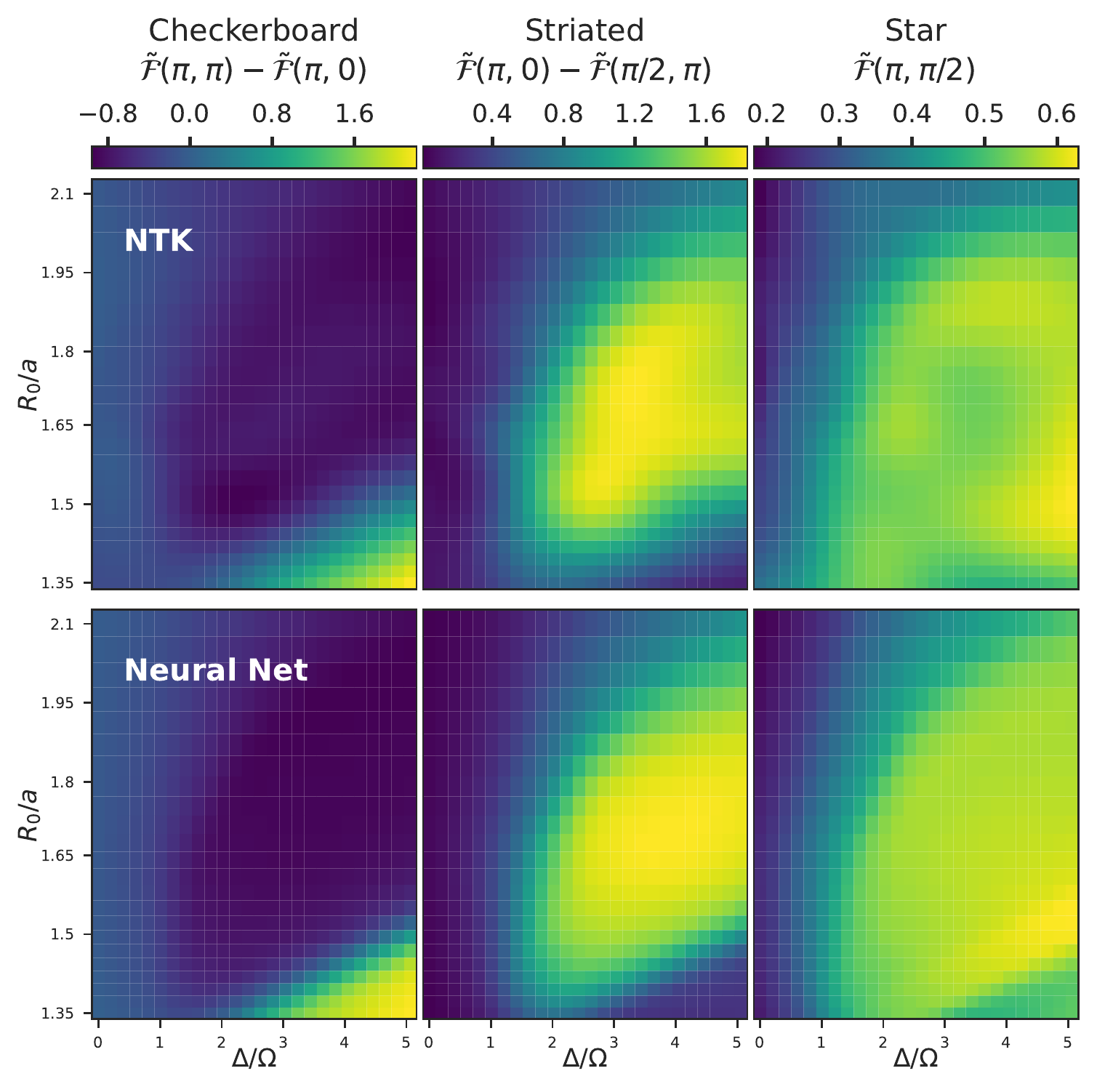}
    \caption{Phase diagrams of 13$\times$13 Rydberg-atom systems predicted by approaches of NTK and neural net.}
    \label{fig:phase-diagrams:ntk-nn}
\end{figure}

\subsubsection{More Details about Sec. \ref{sec:rydberg:proof-of-concept}}\label{apx:exp-details:rydberg:order-params}

~

\paragraph*{\textbf{1D Lattices.}}
Inspired by Ref.~\cite{Rydberg1D-51atoms}, we define the order parameters for the $Z_2$ and $Z_3$ ordered phases as $O_{Z_2}$ and $O_{Z_3}$
\begin{align}
    O_{Z_2} &= \frac{1}{n}\sum_{i=1,3,5,\dots} N_i\\
    O_{Z_3} &= \frac{1}{n}\sum_{i=1,4,7,\dots} N_i
\end{align}
where $n$ is the number of atoms.

The design of these order parameters is quite intuitive. For example, consider the 7-atom 1D lattice shown in Fig. \ref{fig:rydberg-demo}: in the $Z_2$ ordered phase, the expectation value of $O_{Z_2}$ should be very high (upper bounded by 1), since atoms at sites $1,3,5,\dots$ are excited with high probability; in the $Z_3$ ordered phase, the expectation value of $O_{Z_3}$ should be very high (upper bounded by 1), since atoms at sites $1,4,7,\dots$ are excited with high probability.

Similar to the practice in Ref.~\cite{huang2021provably}, we set thresholds on the values of these order parameters to determine the phase of a state: If $\langle O_{Z_2}\rangle > 0.7$, we determine this state as in the $Z_2$ ordered phase; if $\langle O_{Z_2}\rangle  > 0.6$, we determine the state as in the $Z_3$ ordered phase; if a state is not in one of these phases, we classify it as in the disordered phase.

~

\paragraph*{\textbf{2D Square Lattices.}}
The order parameters of the Checkboard and Striated phases are defined in \eqref{eq:order-param:checkboard} and \eqref{eq:order-param:striated}.

For the Staggered phase, \cite{Rydberg2D-256atoms} has not defined its corresponding order parameter, so we define it by ourselves based on our observation of its patterns in the Fourier space:
\begin{align}\label{eq:order-param:staggered}
    O_{\mathrm{staggered}} &= \left (\cF(\pi/2, \pi/2) + \cF(-\pi/2, \pi/2) + \cF(\pi/2, -\pi/2)  +  \cF(-\pi/2,-\pi/2) \right) / 4 
\end{align}

For each prepared state, we apply $\{O_{\textrm{checkboard}}, O_{\textrm{striated}}, O_{\textrm{staggered}}\}$ to it over measurement outcomes to estimate the values of these order parameters, $\{\langle O_{\textrm{checkboard}} \rangle, \langle O_{\textrm{striated}} \rangle, \langle O_{\textrm{staggered}}\rangle\}$. Then, we use a threshold 0.65 to determine the phase: if one of $\{\langle O_{\textrm{checkboard}}/1.6 \rangle, \langle O_{\textrm{striated}}/0.8 \rangle, \langle O_{\textrm{staggered}}\rangle\}$ by is above this threshold (1.6 and 0.8 are normalization factors to rescale the order parameters to the range of $[0,1]$), we determine the state as in the corresponding phase; if all values are below the threshold, we determine the state as in the disordered phase; otherwise, if multiple order parameters are above the threshold (very rare situation), we use the one with the largest value to determine the phase.

\begin{figure*}[ht!]
    \centering
    \subfloat[Checkboard Phase]{\label{fig:2d-rydberg-fourier:checkboard}\includegraphics[width=.31\linewidth]{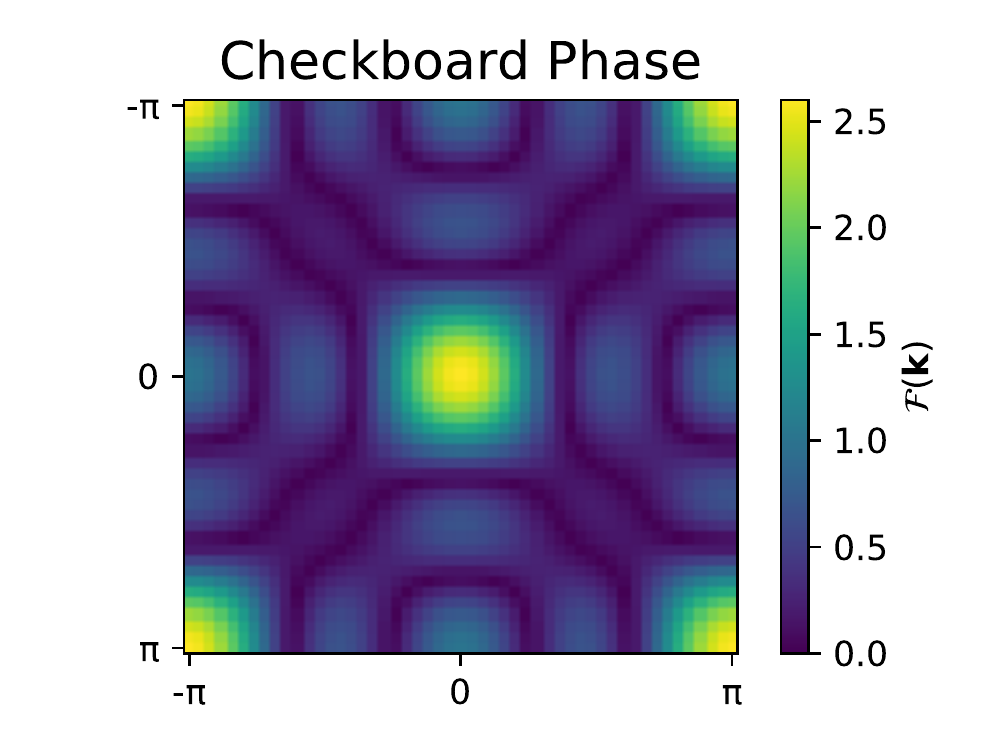}}%
    \hfill
    \subfloat[Striated Phase]{\label{fig:2d-rydberg-fourier:striated}\includegraphics[width=.31\linewidth]{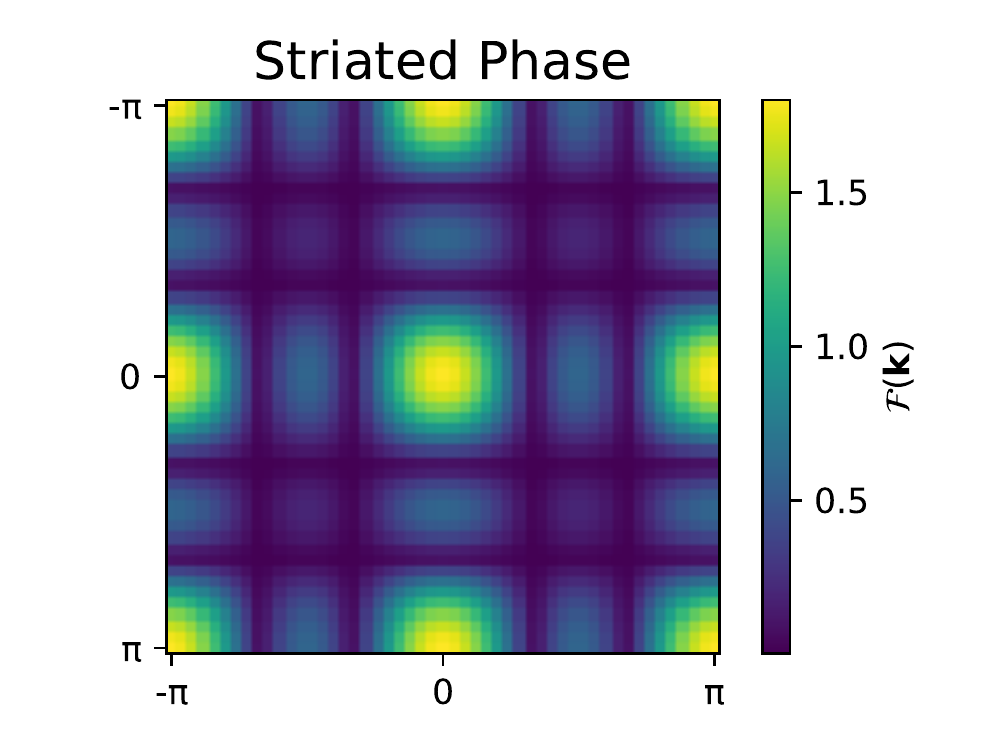}}%
    \hfill
    \subfloat[Staggered Phase]{\label{fig:2d-rydberg-fourier:staggered}\includegraphics[width=.31\linewidth]{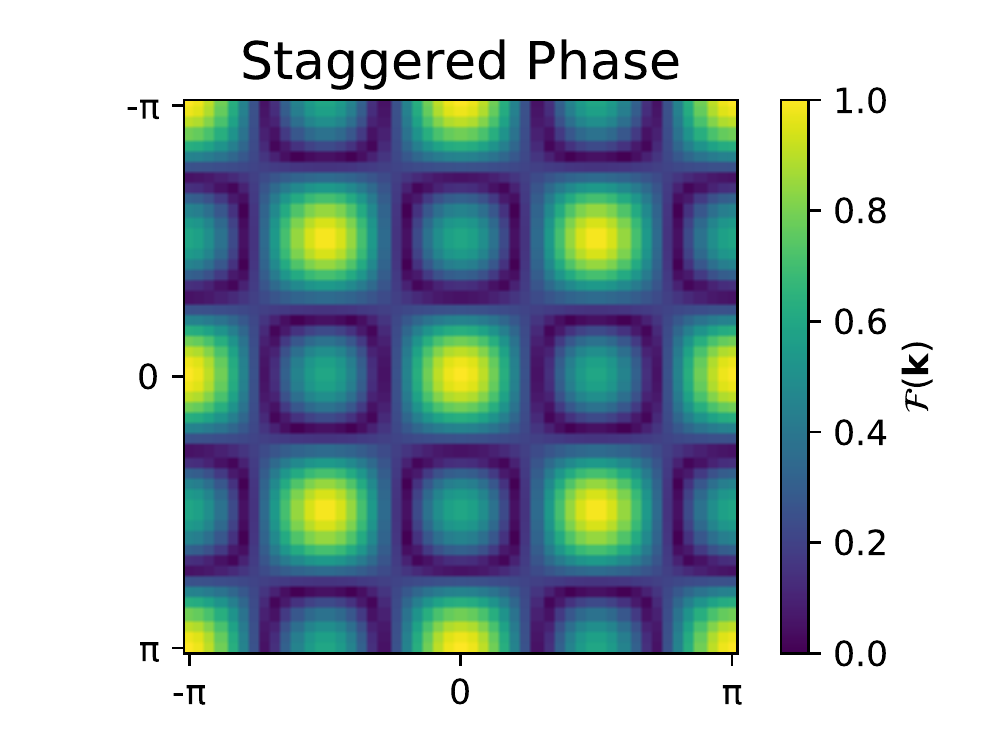}}
    \caption{The Fourier space visualization of square lattices of $5\times 5$ Rydberg atoms in different ordered phases (see Fig. \ref{fig:rydberg-demo} for illustration of these phases.}\label{fig:2d-rydberg-fourier}
\end{figure*}

In Fig. \ref{fig:2d-rydberg-fourier}, we provide a visualization of the Fourier spectra of the three ordered phases. One can see each Fourier spectrum has some dominant peaks, which can be viewed as the signatures of each phase. In fact, the design of order parameters in Ref.~\cite{Rydberg2D-256atoms} is based on this intuition. Notably, since our numerical study is conducted on 5x5 square lattices, the Staggered phase is a little different from the observation in Ref.~\cite{Rydberg2D-numerical} due to finite-system-size effects, so our Fourier space visualization of the Staggered phase is also different from that of Ref.~\cite{Rydberg2D-numerical}. Hence, when studying 2D square lattices beyond the size of 5x5, one should modify our choice of $O_{\textrm{staggered}}$ in \eqref{eq:order-param:staggered} correspondingly.

\section{Classical Simulation}\label{apx:simulation}

\subsection{2D Random Heisenberg Models}\label{apx:simulation:heisenberg}
For 2D random Heisenberg systems with fewer than 20 qubits, we generate 100 random Hamiltonians by sampling coupling constants $J_{ij}\sim\cU[0,\,2]$, and determine their ground states via exact diagonalization. After obtaining the ground state wavefunction, we apply Pauli-6 POVM measurements to it repeatedly following the classical shadow protocol \cite{huang2020predicting}, which effectively corresponds to measuring the state in random Pauli bases. For 2D random Heisenberg systems with 20 qubits or more, we use the publicly available dataset provided by the authors of Ref.~\cite{huang2021provably} (\url{https://github.com/hsinyuan-huang/provable-ml-quantum}), where the ground states are approximated by DMRG using the ITensor package\cite{itensor}. Our implementation is written in Python and partially built on the PennyLane library \cite{pennylane}.

\subsubsection*{Pauli-6 POVM}
The single-qubit Pauli-6 POVM has six outcomes corresponding to sub-normalized rank-1 projections
\begin{equation}
    \cM_{\text{Pauli-}6} = \left\{
        \frac{1}{3}\ketbra{+}{+},\,
        \frac{1}{3}\ketbra{-}{-},\,
        \frac{1}{3}\ketbra{+i}{+i},\,
        \frac{1}{3}\ketbra{-i}{-i},\,
        \frac{1}{3}\ketbra{0}{0},\,
        \frac{1}{3}\ketbra{1}{1}
    \right\}
\end{equation}
where $\{\ket{+},\,\ket{-}\}$, $\{\ket{+i},\,\ket{-i}\}$ and $\{\ket{0},\,\ket{1}\}$ are the eigenbases of the Pauli operators $X$, $Y$ and $Z$, respectively. Since each Pauli matrix and, in addition, the identity matrix, can be obtained from real linear combinations of the projections in $\cM_{\text{Pauli-}6}$, it follows that the single-qubit Pauli-6 POVM spans the space of $2\times2$ Hermitian matrices. The Pauli-6 POVM on $n$ qubits is formed by taking $n$-fold tensor products of the POVM elements in $\cM_{\text{Pauli-}6}$ and is hence informationally complete.

\subsection{Rydberg Atom Systems}\label{apx:simulation:rydberg}

\begin{figure}[ht!]
    \centering
    \includegraphics[width=.8\linewidth]{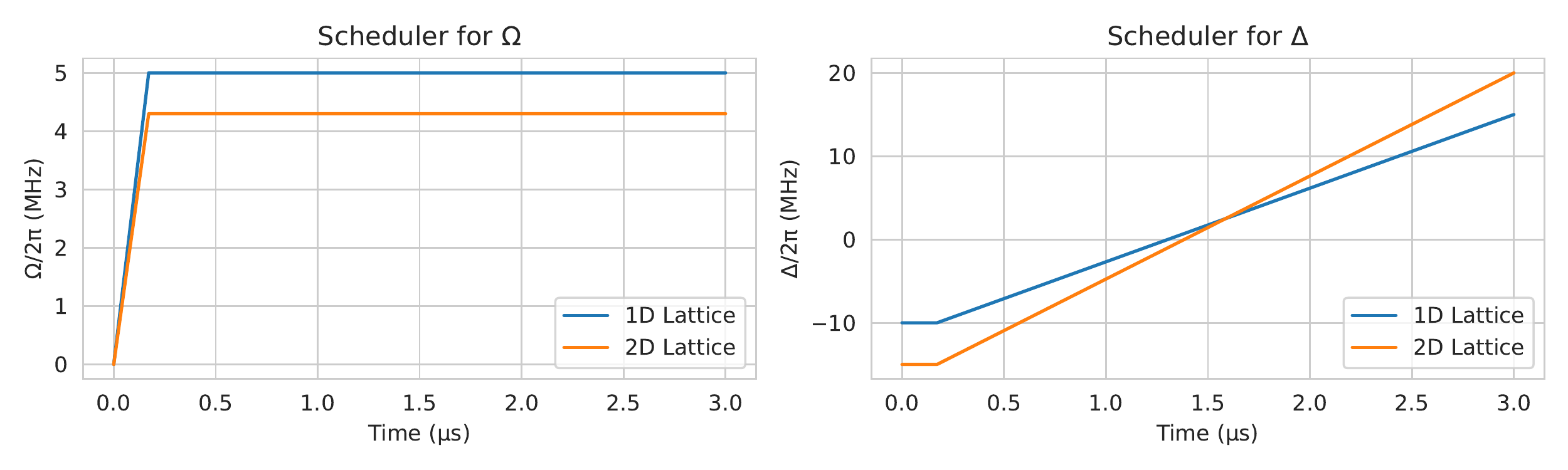}
    \caption{Our schedulers for $\Omega$ and $\Delta$ during an adiabatic evolution of $3\mu s$, for 1D and 2D lattices of Rydberg atoms. For longer or shorter adiabatic evolution times, we just proportionally enlarge or shrink these schedules.}
    \label{fig:rydberg:schedulers}
\end{figure}

Our classical simulations of Rydberg atom systems are conducted via the Bloqade.jl \cite{Bloqade}, a Julia \cite{julia} package, with the following steps:
\begin{enumerate}
    \item We construct a 1D or 2D lattice of Rydberg atoms with a specified lattice configuration (e.g., lattice dimensions and atom separation $a$).
    \item We design schedulers for the adiabatic evolution of $\Omega$ and $\Delta$ (defined in \eqref{eq:H-rydberg}), respectively. We visualize the schedulers we used in Fig. \ref{fig:rydberg:schedulers}.
    \item We start the simulation of adiabatic evolution via an ODE solver. At some preset time steps, we pause the simulation, and apply $Z$-basis measurements to this lattice. Then, we record the measurement outcomes along with the current values of $\Omega$ and $\Delta$, and resume the simulation process.
\end{enumerate}

\textbf{Remarks.}~ Bloqade.jl\cite{Bloqade} considers neutral ${}^{87}$Rb atoms as the Rydberg atoms, with the $70S_{1/2}$ Rydberg state of Rb atoms as the excited state $\ket{r}$. Then, the Rydberg interaction constant becomes $V_0= 862690 \times 2\pi \textrm{ MHz } \mu m^ 
6$ ($V_0$ is also often referred to as $C_6$ in the literature).

\end{document}

%% file: math_commands.tex
\newcommand{\cD}{\mathcal{D}}

\newcommand{\cF}{\mathcal{F}}

\newcommand{\cH}{\mathcal{H}}

\newcommand{\cL}{\mathcal{L}}
\newcommand{\cM}{\mathcal{M}}

\newcommand{\cS}{\mathcal{S}}

\newcommand{\cU}{\mathcal{U}}

\newcommand{\R}{\mathbb{R}}

\newcommand{\Id}{\mathds{1}}

\newcommand{\E}[2][]{ \ifthenelse{\isempty{#1}}
  {\mathbf{\mathbb{E}}\left[#2\right]}
  {\mathbf{\mathbb{E}}_{#1}\left[#2\right]} }

\newcommand{\bx}{\mathbf{x}}

\newcommand{\tr}[1]{\mathrm{Tr}[#1]}

\newcommand{\ket}[1]{\lvert#1\rangle}

\newcommand{\ketbra}[2]{\lvert#1\rangle\langle#2\rvert}

\newcommand{\norm}[2]{\left\|#1\right\|_{#2}}
\makeatletter
\def\norm{\@ifnextchar[{\@with}{\@without}}
\def\@with[#1]#2{\left\|#2\right\|_{#1}}
\def\@without#1{\left\|#1\right\|}
\makeatother

\theoremstyle{plain}